\documentclass[runningheads]{llncs}

\usepackage{booktabs,comment}
\usepackage{colortbl}
\usepackage[ruled]{algorithm2e}
\usepackage[mathscr]{eucal}
\usepackage{tikz}
\usetikzlibrary{arrows, automata,shapes}
\usetikzlibrary{decorations.pathreplacing}
\usepackage{hyperref,multirow,xcolor}
\usepackage{comment,amsmath,amssymb,amstext,amsfonts,graphics, enumerate}
\usepackage{color,graphicx,latexsym,amsfonts,epsfig,eufrak,array,boxedminipage,graphicx,bm}

\SetAlFnt{\small}
\SetAlCapFnt{\small}
\SetAlCapNameFnt{\small}
\SetAlCapHSkip{0pt}
\IncMargin{-\parindent}

\newcommand \dia{\hfill{\textcolor{darkgray}{$\diamond$}}}

\newcommand{\R}{\mathbb{R}}
\newcommand{\NP}{{\sf NP}}

\begin{document}
\title{Computing Balanced Solutions for Large International Kidney Exchange Schemes\thanks{Benedek acknowledges the support of the National Research, Development and Innovation Office of Hungary (OTKA Grant No.\ K138945). Bir\'o acknowledges the support of the Hungarian Scientific Research Fund (OTKA, Grant No.\ K129086). Paulusma acknowledges the support of the Leverhulme Trust (Grant RF-2022-607).}}

\author{M\'arton Benedek\inst{1,2} \and
P\'eter Bir\'o\inst{1,2} \and
Daniel Paulusma\inst{3}
\and
Xin Ye\inst{4}}

\authorrunning{M. Benedek et al.}
\institute{KRTK, Hungary \email{\{peter.biro,marton.benedek\}@krtk.hu}\and
Durham University \email{\{daniel.paulusma,xin.ye\}@durham.ac.uk}}

\maketitle

\begin{abstract}
\vspace*{-3mm}
To overcome incompatibility issues, kidney patients may swap their donors.
In international kidney exchange programmes (IKEPs), countries merge their national patient-donor pools.
We consider a recently introduced credit system. In each round, countries are given an 
initial ``fair'' allocation of the total number of kidney transplants. 
This allocation is adjusted by a credit function yielding a target allocation. The goal is to find a solution that approaches the target allocation as closely as possible, to ensure long-term stability of the international pool. As solutions, we use maximum matchings that
lexicographically minimize the country deviations from the target allocation. We perform, for the first time, a computational study for a {\it large} number of countries.
For the initial allocations we use two easy-to-compute solution concepts, the benefit value and the contribution value, and four classical but hard-to-compute concepts, the Shapley value, nucleolus, Banzhaf value and tau value. 
By using state-of-the-art software we show that the latter four concepts are now within reach for IKEPs of up to fifteen countries.
Our experiments show that
using lexicographically minimal maximum matchings
instead of ones that only minimize the largest deviation from the target allocation (as previously done)
may make an IKEP up to 54\% more balanced. 
\end{abstract}

\section{Introduction}\label{s-intro}

For kidney patients, kidney transplantation is still the most effective treatment resulting in a significant longer 
life expectancy compared to dialysis.  However, the demand for available kidneys has consistently exceeded supply. Moreover,
a kidney transplantation might not be possible due to blood-type or tissue-type incompatibilities between a patient and a willing donor. The solution is to place all patient-donor pairs in one pool such that donors can be swapped in a cyclic manner. More formally, an {\it $\ell$-way exchange} involves $\ell$ distinct patient-donor pairs $(p_1,d_1),\ldots, (p_\ell,d_\ell)$, where for $i\in \{1,\ldots,\ell-1\}$, donor $d_i$ donates to patient $p_{i+1}$ and donor $d_{\ell}$ donates to patient $p_1$. A {\it kidney exchange programme} (KEP) is a centralized program where the goal is to find an optimal kidney exchange scheme in a pool of patient-donor pairs subject to an upper bound $\ell$ on the cycle length.

Recently, national KEPs started to collaborate, leading to a number of {\it international} KEPs 
(IKEPs). 
In 2016, the first international kidney exchange took place, between Austria and the Czech Republic~\cite{Bo_etal17}. In 2018, Italy, Spain and Portugal started to collaborate~\cite{Va_etal19}. In 2019, Scandiatransplant, an organization for sharing deceased organs among six Scandinavian countries, started an IKEP involving Swedish and Danish transplant centers.
Even though overall solutions will improve, {\it individual rationality} might not be guaranteed, that is, individual countries could be worse off.
To improve the stability of an IKEP,  the following question is therefore highly relevant:

\medskip
\noindent
{\it What kind of fairness must we guarantee to ensure that all countries in an IKEP place all their patient-donor pairs in an international pool?}

\medskip
\noindent
Individual rationality~\cite{AR12,AR14} and fairness versus optimality~\cite{As_etal15,Blum_etal2017,Ha_etal15,TP15} were initially studied for national KEPs, in particular in the US.
  However, the US situation is
different from
many other countries.
The US has three nationwide KEPs (UNOS, APD, NKR)~\cite{Ag_etal18}, and
US hospitals work independently and compete with each other. Hence, US hospitals tend to register only the hard-to-match patient-donor pairs to the national KEPs, while they try to process their easy-to-match pairs immediately. As a consequence, the aforementioned papers focused on mechanisms that give incentives for hospitals to register all their patient-donor pairs at the KEP. In particular, NKR (the largest nationwide KEP in the US) uses a credit system to incentivize hospitals to register also their easy-to-match pairs by giving negative credits for registering hard-to-match pairs and positive credits for registering easy-to-match pairs.

\subsection{Our Setting}
We consider IKEPs in the setting of European KEPs which are scheduled in rounds,
typically once in every three months~\cite{Biro_etal2019}. Unlike the US setting, this setting allows a search for optimal exchange schemes. Hence, the situation where easy-to-match patient-donor pairs are taken out of the pool is no longer relevant.
 Below we discuss existing work for the European setting. As we will see, the credit system proposed for the European setting~\cite{BKPP19,KNPV20} is different from the one used by NKR due to the different nature of the European and US settings.
 
We first note that the search for an optimal exchange scheme can be done in polynomial time for $2$-way exchanges ({\it matchings}) but becomes \NP-hard as soon as $3$-way exchanges are permitted~\cite{Abr07}. 

Carvalho and Lodi~\cite{CL23} used a $2$-round system for ensuring stability of IKEPs with $2$-way exchanges only: in the first round each country selects an internal matching, and in the second round a maximum matching is selected for the international exchanges. They gave a polynomial-time algorithm for computing a Nash-equilibrium that maximizes the total number of transplants, improving the previously  known result of~\cite{Carvalho_etal2017} for two countries.

Sun et al. \cite{STW21} also considered $2$-way exchanges only. They defined so-called selection ratios using various lower and upper target numbers of kidney transplants for each country. In their setting, a solution is considered to be fair if the minimal ratio across all countries is maximized. They also required a solution to be a maximum matching and individually rational.
They gave theoretical results specifying which models admit solutions with all three properties. Moreover, they provided polynomial-time algorithms for computing such solutions, if they exist.

Klimentova et al.~\cite{KNPV20} introduced a credit system to incentivize the countries for collaborating by sharing the joint benefits in a fair way. That is, each country will be allocated in each round of the IKEP a ``fair'' target number of kidney transplants for that country. The differences between the actual number of transplants for a country and its target number are used as {\it credits} for the next round. In their simulations, they allowed $3$-way exchanges for four countries. They considered the potential and benefit value for the initial allocations, which become the target allocations after the credit adjustment.
Their results showed that using the benefit value yields slightly more balanced solutions. Bir\'o et al.~\cite{BGKPPV20} compared the benefit value with the Shapley value. In their simulations, for three countries allowing $3$-way exchanges, they found that the Shapley value produced smaller deviations from the targets on average. 

Bir\'o et al.~\cite{BKPP19} considered credit-based compensation systems from a theoretical point of view. They only allowed for $2$-way exchanges but, unlike~\cite{BGKPPV20,KNPV20}, with the possibility of having weights for representing transplant utilities. They gave a polynomial-time algorithm for finding a maximum matching that minimizes the largest country deviation from a target allocation. They also showed that the introduction of weights makes the problem \NP-hard.
In~\cite{BBKP22}, the polynomial-time algorithm~of~\cite{BKPP19} was generalized to a polynomial-time algorithm for computing a maximum matching that {\it lexicographically} minimizes the country deviations from a given target allocation. In~\cite{BBKPP}, the theoretical results from~\cite{BBKP22} and~\cite{BKPP19} are unified and extended.

\subsection{Our Contributions}
We perform a {\it large scale} experimental study (up to 15 countries) on finding balanced solutions in IKEPs. 
Our motivation is threefold. 
Firstly, these days, IKEPs have a growing number of countries. Secondly, we aim to measure the effect of using maximum matchings that are lexicographically minimal. We therefore need to consider IKEPs with a large number of participating countries (otherwise maximum matchings that minimize the largest country deviation from a target allocation will probably be lexicographically minimal). 
Thirdly, motivated by the promising results for the Shapley value~\cite{BGKPPV20,KNPV20}, we also wanted to thoroughly investigate the effect of using widely accepted solution concepts from cooperative game theory. That is, we model the rounds of an IKEP as so-called partitioned matching games, which were formally introduced in~\cite{BKPP19}. This indeed allows us to use well-understood solution concepts from cooperative game theory for prescribing the initial ``fair'' allocations. We define all game-theoretic notions that we need in Section~\ref{s-pre}. 

In Section~\ref{s-model} we explain the credit system of~\cite{BBKP22,BGKPPV20,BKPP19,KNPV20} in our setting. 
Whilst~\cite{BGKPPV20,KNPV20} allowed $3$-way exchanges, we only allow $2$-way exchanges just as \cite{Carvalho_etal2017,CL23,STW21}.
Similar to~\cite{BGKPPV20,KNPV20} we do not consider weights representing transplant utilities. We justify our setting as follows.
We first recall that allowing $3$-way exchanges~\cite{Abr07} or weights~\cite{BKPP19} makes the problem of computing an optimal exchange scheme in a certain round  \NP-hard. With the current technology it is not possible to perform an experimental study on such a large scale as we do. Furthermore, some countries, such as France and Hungary, are legally bound to using only $2$-way exchanges. Hence, assuming only $2$-way exchanges is not unrealistic either. Moreover, in most of the existing KEPs, the primary objective does not involve any weights and is still to maximize the number of kidney 
transplants~\cite{Biro_etal2021}.

In Section~\ref{s-poly} we describe the algorithm, called {\tt Lex-Min}, that we used for computing maximum matchings that lexicographically minimize the country deviations from a given target allocation. As we will explain, this algorithm can also be used for computing maximum matchings that minimize only the largest country deviation from a given target allocation. For the correctness proof and a running time analysis of the algorithm we refer to~\cite{BBKP22} (see also~\cite{BBKPP}).

In Section~\ref{s-sim} we discuss the simulations in more detail, and in Section~\ref{s-results} we present the results of our simulations. As mentioned, we conduct simulations 
for up to even 15 countries in contrast to the previous studies~\cite{BGKPPV20,KNPV20} for 3--4 countries. Moreover, we do this both for equal and varying country sizes, and for a large variety of different solution concepts. 
Namely, our target allocations are prescribed by four hard-to-compute solution concepts: the Shapley value, nucleolus, Banzhaf value and tau 
value.\footnote{In 0.04\% of our simulations, the tau value does not exist and in those cases we replace the tau value by the closely related benefit value.}
and two easy-to-compute solution concepts: the aforementioned benefit value, 
which coincides with the Gately point if the latter is unique,
 and a natural variant of the benefit value, the contribution 
value.\footnote{Our experimental study is based on the one in~\cite{BBKP22}, but we made a significant extension: we now also included the Banzhaf value and tau value in our simulations.} As mentioned, we define all these concepts in Section~\ref{s-pre}.

Our simulations show that a credit system using lexicographically minimal maximum matchings instead of ones that minimize only the largest country deviation from a given target allocation makes an IKEP up to 54\% more balanced,  without decreasing the overall number of transplants. The exact improvement depends on which solution concept is used. From our experiments, both the Banzhaf and Shapley value yield the best results, namely, on average, a deviation of up to 0.52\% from the target allocation. However, the differences between the different solution concepts are small: all the other solution concepts stay within 1.23\% from the target allocation, and the choice for using a certain solution concept will be up to the policy makers of the IKEP.

We finish our simulations by examining a new approach for incorporating credits that has not been proposed in the literature before. Namely, it is also natural to let the solution concepts prescribe an allocation for a {\it credit-adjusted} game, where the credits are incorporated into the value function of the game directly. 
As explained in Section~\ref{s-model}, where we introduce this approach after first describing the original model, only the Banzhaf value may prescribe different allocations. For all the other solution concepts that we consider both the original and new credit system yield exactly the same target allocations.
Our simulations show, however, that the modified Banzhaf value yields, on average, a deviation of even up to 0.48\% from the target allocation. This is a slight improvement over the best results (0.52\%) under the original credit system.

In Section~\ref{s-evaluation} we evaluate some other aspects of our simulations. First, we show that IKEPs lead to a significantly larger number of total kidney transplants than the total number of transplants of the KEPs of the individual countries. Second, we show that, although theoretically country credits may build up over time as illustrated with an example in Section~\ref{s-model}, this situation does not happen in any of our simulations. Third, we evaluation computational time issues in our simulations, showing that the generation of the partitioned matching games is the most expensive operation in our simulations.
Fourth, we evaluate a number of game-theoretic properties: core stability aspects, convexity and quasibalancedness.
Finally, in Section~\ref{s-con}, we give directions for future work.

\section{Cooperative Game Theory}\label{s-pre}

We model rounds of IKEPs as partitioned matching games~\cite{BKPP19}.
In this section we define these games. We first provide relevant definitions from cooperative game theory that we will need in the remainder of our paper.

A \emph{(cooperative) game} is a pair $(N,v)$, where $N$ is a set $\{1,\ldots,n\}$ of \emph{players} and $v: 2^N\to \R_+$  is a \emph{value function} with $v(\emptyset) = 0$. A {\it coalition} is a subset $S\subseteq N$.  If $v(N)$ is maximum over all partitions of $N$, then the players have an incentive to form the {\it grand coalition}~$N$.
A {\it partitioned matching game}~\cite{BKPP19} is a game $(N,v)$ on an undirected graph~$D=(V,E)$ with a positive edge weighting~$w$ and a partition $(V_1,\ldots,V_n)$ of $V$. For $S \subseteq N$, we let $V(S)=\bigcup_{p \in S} V_p$. 
A {\it matching}~$M$ in  a graph is a set of edges, such that no two edges in $M$ have an end-vertex in common. The weight of $M$ is $w(M)=\sum_{e\in M}w(e)$. 
The value $v(S)$ of coalition~$S$ is the maximum weight of a matching in the subgraph of $D$ induced by $V(S)$.
If $V_p=\{p\}$ for $p=1,\ldots,n$, then we obtain the classical {\it matching game} (see, for example,~\cite{BKP12,DIN99,IKKKO16,KP03,KPT20}).
We will mainly consider {\it uniform} matching games, that is, with $w(e)=1$ for every $e\in E$. Now, $v(S)$ becomes the maximum size of a matching in the subgraph of~$D$ induced by $V(S)$, and in particular $v(N)=\mu$, where $\mu$ is the size of a maximum matching in $D$.

The central problem in cooperative game theory is how to distribute $v(N)$ amongst the players in such a way that players are not inclined to leave the grand coalition.
In this context, an {\it allocation} is a vector $x \in \R^N$ with $x(N) = v(N)$
where we write $x(S)=\sum_{p\in S}x_p$ for some set $S\subseteq N$; hence, $x_i$ prescribes the part of $v(N)$ that is allocated to player~$i$. An allocation~$x$ is said to be an {\it imputation} if $x$ is {\it individual rational}, that is, $x_p\geq v(\{p\})$ for every $p\in N$.
A {\it solution concept} prescribes a set of ``fair'' allocations for cooperative games, where the notion of fairness depends on context. We now provide definitions of the solution concepts that are relevant for our work.

The {\it core} of a game $(N,v)$ consists of all allocations $x \in \R^N$ with $x(S)\geq v(S)$ for every $S\subseteq N$.
Core allocations offer no incentive for a group of players to leave the grand coalition~$N$ and form a coalition on their own, so they ensure that $N$ is stable. However, games may have an empty core.

We now define the nucleolus of a cooperative game. In order to do this we need more terminology.
For an allocation~$x$ and a non-empty coalition $S \subsetneq N$ in a game ($N,v$), we define the {\it excess} $e(S,x) := x(S)-v(S)$. We obtain the {\it excess vector} $e(x) \in  \R^{2^n-2}$  by ordering the $2^n-2$ entries in a non-decreasing sequence.
The {\it nucleolus} of a game ($N,v$) is the unique allocation~\cite{Sc69} that lexicographically maximizes $e(x)$ over the set of imputations. Note that the nucleolus is not defined if the set of imputations is empty. However, every partitioned matching game has a nonempty set of imputations. 
The nucleolus is a core allocation if the core is nonempty.

The {\it Shapley value} $\phi(N,v)$ of a game $(N,v)$, defined in~\cite{Sh53}, is one of the best known solution concepts and is defined by setting for every $p\in N$,
\begin{equation}\label{eq-Shapley}
\phi_p(N,v) = \displaystyle\sum_{S \subseteq N\backslash \{p\}}
 \frac{|S|!(n-|S|-1)!}{n!}\bigg(v(S\cup \{p\})-v(S)\bigg).
 \end{equation}
Unlike the nucleolus, the Shapley value does not necessarily belong to the core if the core is nonempty. This also holds for (partitioned) matching games (see~\cite{BBJPY} for a small example).

The {\it unnormalized Banzhaf value} $\psi_p(N,v)$ of a game $(N,v)$ is introduced in~\cite{Ba64} and is defined by setting for every $p\in N$,
\begin{equation}\label{eq-ubv}
\psi_p(N,v):=\displaystyle\sum_{S \subseteq N\backslash \{p\}} \frac{1}{2^{n-1}}\bigg(v(S\cup \{p\})-v(S)\bigg).
 \end{equation}
 Note that $\psi_p$ may not be an allocation. The {\it (normalized) Banzhaf value} $\overline{\psi}_p(N,v)$ of a game $(N,v)$ rectifies this and is defined by setting for every $p\in N$,
 \begin{equation}\label{eq-nbv}
\overline{\psi}_p(N,v):=\displaystyle\frac{\psi_p(N,v)}{\sum_{q \in N}\psi_q(N,v)} \cdot v(N).
 \end{equation}
 Whenever we speak about the Banzhaf value in our paper, we will mean $\overline{\psi}(N,v)$.

We now define the tau value~\cite{T81}. Let $(N,v)$ be a game. For $p\in N$, let $b_p = v(N) - v(N \setminus \{p\})$ be the \emph{utopia payoff} for~$p$. This gives us a vector $b\in \R^n$. For $p\in N$ and $S\subseteq N$, let  $R(S,p):=v(S)-\sum_{q \in S\setminus\{p\}}b_q$ be what remains for $p$ should the other players in $S$ leave $S$ with their utopia payoff. For $p\in N$, we let $a_p:= \max_{S \ni p} R(S,p)$. This gives us a vector $a\in \R^n$. We say that $(N,v)$ is {\it quasibalanced} if both $a\leq b$ (that is, $a_p\leq b_p$ for every $p\in N$) and $a(N) \leq v(N) \leq b(N)$.  For a quasibalanced game $(N,v)$, the {\it tau value}~$\tau$ is defined by setting for every $p\in N$,
$$\tau_p:=\gamma a_p + (1-\gamma)b_p,$$ where $\gamma \in [0,1]$ is determined by the condition $\tau(N) = v(N)$. Note that $\gamma$ is unique unless $a=b$ in which case $\tau=a$. The tau value is not defined for games that are not quasibalanced.

In general, all the above solution concepts may require exponential time to compute, assuming that the input is described by an underlying (weighted) graph, as in the case of partitioned matching games.
We now define two easy-to-compute solution concepts.
To do this, we first define the {\it surplus} of a game $(N,v)$ as $$\mbox{surp}=v(N) - \sum_{p \in N} v(\{p\}).$$ 
A game $(N,v)$ is said to be {\it essential} if $\mbox{surp}>0$. Note that an essential game has more than one imputation. If $\mbox{surp}=0$, then the allocation $x\in \R^n$ with $x_p=v(\{p\})$ for every $p\in N$ is the unique imputation, whereas the set of imputations is empty if $\mbox{surp}<0$. As mentioned, partitioned matching games have a nonempty set of imputations, but they may not be essential; consider, for instance, a matching game 
defined on a graph consisting of two non-adjacent vertices.

For $p\in N$ we can allocate $v(\{p\})+\alpha_p\cdot\mbox{surp}$ for some~$\alpha\in \R^n$ with $\sum_{p\in N}\alpha_p=1$. We define two solution concepts that each correspond to a different $\alpha$.

First, we obtain the known \emph{benefit value}~\cite{KNPV20} by setting for each $p\in N$, $$\alpha_p=\frac{v(N) - v(N \setminus \{p\})-v(\{p\})}{\sum_{p\in N}(v(N) - v(N \setminus \{p\})-v(\{p\}))}.$$ The benefit value is not defined if $\sum_{p\in N}(v(N) - v(N \setminus \{p\})-v(\{p\})) = 0$. 

Moreover, partitioned matching games are {\it superadditive}, that is, for every two disjoint coalitions $S$ and $T$, it holds that $v(S \cup T) \geq v(S) + v(T)$. This means that the benefit value coincides with the Gately point~\cite{Ga74}, as long as the Gately point is unique. For superadditive games, the latter holds if there exists at least one player $p$ with 
 $v(N) - v(N \setminus \{p\})-v(\{p\})>0$~\cite{SA19}. For superadditive games that do not satisfy this condition, we have for every $p\in N$ that $v(N)-v(N \setminus \{p\})-v(\{p\})=0$, and thus the benefit value does not exist. Moreover,
 the benefit value coincides with the tau value when the game is {\it convex}~\cite{Ya10}, that is, for every two coalitions $S$ and $T$ it holds that $v(S \cup T) + v(S \cap T) \geq v(S) + v(T)$. This condition implies superadditivity. However, already any (uniform) matching game that contains a 3-vertex path $uvw$ as a subgraph is not convex: take $S=\{u,v\}$ and $T=\{v,w\}$ and note that $v(S\cup T)+v(S\cap T)=1<2=v(S)+v(T)$.\footnote{The tau value and benefit value may also coincide for non-convex matching games: e.g. take the $4$-vertex cycle with unit edge weights, for which the allocation $x\equiv\frac{1}{2}$ is both the tau and benefit value.}

Second, we obtain a new solution concept, the \emph{contribution value}, by setting for each $p\in N$,
$$\alpha_p=\frac{v(N) - v(N \setminus \{p\})}{\sum_{p\in N}(v(N) - v(N \setminus \{p\}))}.$$ The contribution value is not defined if $\sum_{p\in N}(v(N) - v(N \setminus \{p\})) = 0$. 
From their definitions, we note that both the benefit value and the contribution value can  be computed in polynomial time.

Finally, we observe that there exist even small matching games for which the tau value, benefit value and contribution value do not to exist and for which the Gately point is not unique, while the set of imputations has size larger than~$1$. Namely, let $D$ be the triangle with unit edge weights.

\section{Our Model}\label{s-model}

We model a KEP as follows.
A {\it compatibility graph} is a directed graph~$D=(V,A)$ with an arc weighting $w$. Each vertex in~$V$ is a patient-donor pair. There is an arc from patient-donor pair~$i$ to patient-donor pair~$j$ if and only if the donor of pair~$i$ is compatible with the patient of pair~$j$.
The associated weight~$w_{ij}$ indicates the utility of the transplantation. An {\it  exchange cycle} is a directed cycle~$C$ in~$D$. The {\it weight} of a cycle~$C$ is the sum of the weights of its arcs.  An {\it  exchange scheme}~$X$ is a union of pairwise vertex-disjoint exchange cycles in~$D$. The weight of $X$ is the sum of the weights of its cycles.

A KEP operates in rounds. Each round has its own compatibility graph, which is determined by the current pool of patient-donor pairs. In each round,
the goal is to find an exchange scheme of maximum weight, subject to a fixed {\it exchange bound} $\ell$, which is an upper bound on the length of the exchange cycles that may be used.

We obtain an IKEP by partitioning $V$ into
subsets $V_1,\ldots,V_n$, where $n$ is the number of countries involved and for $p\in \{1,\ldots,n\}$, $V_p$ is the set of patient-donor pairs of country~$p$. The objective is still to find an exchange scheme of $D$ that has maximum weight subject to the exchange bound~$\ell$. In this setting, we must now in addition ensure 
that the countries accept the proposed exchange schemes.

\begin{figure}
\begin{center}
\begin{tabular}{cc}
 \begin{tikzpicture}[
            > = stealth,
            shorten > = 1pt,
            auto,
            node distance = 3cm,
            semithick
        ]

        \tikzstyle{every state}=[scale=0.55,
            draw = black,
            thick,
            fill = white,
            minimum size = 4mm
        ]
        \node[state] (i2) {$i_2$};
        \node[state] (i1) [above right of=i2] {$i_1$};
        \node[state] (j2) [right of=i2] {$j_2$};
        \node[state] (j1) [below right of=i2] {$j_1$};
        \node[state] (i3) [right of=j2] {$i_3$};
 \path[->]  (i1) edge node {} (i2);
        \path[->] [bend left] (i2) edge node {} (i1);
        \path[->]  (i2) edge node {} (j2);
        \path[->] [bend left] (j2) edge node {} (i2);
        \path[->] (i2) [bend right] edge node {} (j1);
        \path[->] (j2) edge node {} (i1);
                \path[->] (j2) [bend left] edge node {} (j1);
        \path[->] (j1) edge node {} (j2);
        \path[->] (i1) edge node {} (i3);
    \end{tikzpicture}
    \hspace*{1.5cm}
     \begin{tikzpicture}[
            > = stealth,
            shorten > = 1pt,
            auto,
            node distance = 3cm,
            semithick
        ]

        \tikzstyle{every state}=[scale=0.55,
            draw = black,
            thick,
            fill = white,
            minimum size = 4mm
        ]

        \node[state] (i2) {$i_2$};
        \node[state] (i1) [above right of=i2] {$i_1$};
        \node[state] (j2) [right of=i2] {$j_2$};
        \node[state] (j1) [below right of=i2] {$j_1$};
        \node[state] (i3) [right of=j2] {$i_3$};
 \path[]  (i1) edge node {} (i2);
        \path[]  (i2) edge node {} (j2);
        \path[] (j1) edge node {} (j2);
    \end{tikzpicture}
\end{tabular}
\end{center}
\caption{\label{fig:particular-graphs}A directed compatibility graph (left) which we make undirected (right). Here, ${\cal M}=\{M\}$, where $M=\{i_1i_2,j_1j_2\}$. If $V_1=\{i_1,i_2,i_3\}$ and $V_2=\{j_1,j_2\}$, then $s_1(M)=s_2(M)=2$. That is, both countries receive two kidney transplants if maximum matching $M$ is used, so all transplants are ``in-house''.}\label{f-ex}
\end{figure}
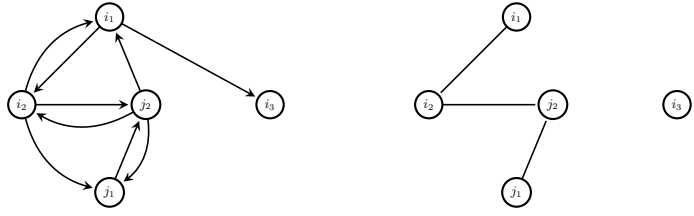

\smallskip
\noindent
{\bf Assumptions.} As explained in Section~\ref{s-intro}, we set  $\ell=2$ and $w\equiv 1$.
As $\ell=2$, we can make $D=(V,A)$ undirected by adding an edge between two vertices $i$ and $j$ if and only if both $(i,j)$ and $(j,i)$ are in $A$ (see Figure~\ref{f-ex}).
So, from now on, compatibility graphs are undirected graphs.

\noindent
We now explain the recent credit system from~\cite{BKPP19,KNPV20} for IKEPs.
Let $N=\{1,\ldots,n\}$ be the set of countries. For some $h\geq 1$, let $D_h$ be the compatibility graph in round~$h$ with country vertex sets $V_1^h,\ldots, V_n^h$.
Let $\mu_h$ be the size of a maximum matching in $D_h$, so $2\mu_h$ is the maximum number of kidney transplants possible in round~$h$. Hence,
an {\it allocation} for round~$h$ is a vector $x^h \in \R^N$ with $x^h(N) = 2\mu_h$. That is, $x^h_p$ describes the share of $x^h(N)=2\mu_h$ that is allocated to country~$p$.
We can only allocate integer numbers (kidneys), but nevertheless we do allow allocations $x^h$ to be non-integer, as we will explain later.

Assume that we are given a ``fair'' allocation~$y^h$ for round $h\geq 1$, together with a {\it credit function} $c^h:N\to \R$, which satisfies $$\sum_{p\in N}c^h_p=0.$$ We let $c^1\equiv 0$ and define $c^h$ for $h\geq 2$ below.
For $p=1,\ldots,n$, we set $x^h_p=y^h_p+c^h_p.$ Then $x^h$ is an allocation, as $y^h$ is an allocation and $\sum_{p\in N}c_p^h=0$. We call $x^h$ the {\it target allocation} for round~$h$ and $y^h$ the {\it initial allocation} for round~$h$.

We now define $c^h$ for $h\geq 2$.
Let ${\cal M}^h$ be the set of all maximum matchings of~$D_h$. Say, we choose a matching $M^h\in {\cal M}^h$.
Then the set $\{(i,j)\in E |\; ij\in M^h, j\in V_p^h\}$ consists of all kidney transplants in round~$h$ that involve patients in country~$p$ (with donors both from country~$p$ and other countries).
We let $s_p(M^h)$ denote the size of this set, or equivalently (see Figure~\ref{f-ex}),
 $$s_p(M^h)=|\{j\in V_p^h |\; ij\in M^h\}|.$$ 
 
 \smallskip
 \noindent
 We now compute a new credit function $c^{h+1}$ by setting $c^{h+1}_p=x^h_p-s_p(M^h)$ and note that $\sum_{p\in N}c_p^{h+1}=0$, as required.
For round~$h+1$, a new initial allocation~$y^{h+1}$ is given. For $p=1,\ldots,n$, we set $x_p^{h+1}=y_p^{h+1}+c_p^{h+1}$ and repeat the process. 

Note that for every country $p\in N$ and round $h\geq 2$, it holds that $$c^h_p=\sum_{t=1}^{h-1}(y_p^t - s_p(M^t)),$$ so credits are in fact the
accumulation of the deviations from the initial allocations. Hence, credits for a country can build up over time, irrespectively of our choice of initial allocations. Later in this section, we will give an explicit example where this happens. However, such situations did not occur in any of our simulations where we used the credit function 
(see Section~\ref{s-ca}).

We now specify our choices for the initial allocations~$y^h$ and maximum matchings $M^h\in {\cal M}^h$.

\medskip
\noindent
{\bf Choosing the initial allocation ${\mathbf y}$.}\\
For prescribing our initial allocations we use the singleton solution concepts from Section~\ref{s-pre}. That is, we use four hard-to-compute solution concepts: the Shapley value, nucleolus, Banzhaf value and tau value, and two easy-to-compute solution concepts: the  benefit value and the contribution value.

We use the same solution concept consistently for all rounds but with {\it one exception}. 
Recall that a partitioned matching game may not be quasibalanced, and that in that case the tau value is not defined. If the game is not quasibalanced, then for our simulations involving the tau value, we use the benefit value instead.\footnote{The benefit value, and also the contribution value, may not always exist either. We kept track of this in our simulations, but both allocations  turned out to existed for all our simulation instances.}
As we will see in Section~\ref{s-sim}, where we describe our simulations in more detail, we had to make this replacement in only 0.04\% of our simulations. 

Naturally, we could have replaced the tau value by a different solution concept. However, we chose the benefit value, as for convex games the tau and benefit values coincide. Moreover, they may even coincide if the game is not convex. Indeed, as we will see in Section~\ref{s-convex}, overall only 4.14\% of the partitioned matching games in our simulations turned out to be convex, but in 31.6\% of the non-convex cases, the tau and benefit values still coincided.

Finally, recall that $x(N)=2\mu$ for an allocation $x$, as we count the number of kidney transplants instead of using the maximum number $\mu$ of patient-donor swaps. To resolve this incompatibility, we multiply the allocations prescribed by the above six solution concepts by a factor of two.

\medskip
\noindent
{\bf Choosing the solution} ${\bm M}${\bf .}\\
For a maximum matching $M\in {\cal M}$ in a partitioned matching game $(N,v)$ we let 
$$\delta(M)= (|x_{p_1}-s_{p_1}(M)|, \dots, |x_{p_n}-s_{p_n}(M)|)$$ 
be the vector obtained by reordering the components $|x_p-s_p(M)|$ non-increasingly. We say that $M$
is {\it lexicographically minimal} for an allocation~$x$ if $\delta(M)$ is lexicographically minimal over all matchings $M\in {\cal M}$. Every lexicographically minimal matching in ${\cal M}$ is a matching that minimizes $$d_1=\max_{p\in N} \{|x_p-s_p(M)|\},$$ but the reverse might not be true. 

In our simulations, we choose in each round a maximum matching that is lexicographically minimal for the target allocation. To examine the effect of this, we will perform exactly the same simulations when we choose a maximum matching that only minimizes $d_1$.
In Section~\ref{s-poly} we present the polynomial-time algorithm for computing these maximum matchings.
Moreover, as a baseline approach, we will do the same simulations when an arbitrary maximum matching is chosen.

\begin{figure}[t]
\begin{center}
\begin{tabular}{cc}
  \begin{tikzpicture}[-, >=stealth',scale = 0.15, inner sep=2mm]
\tikzstyle{nd}=[circle,draw,fill=black!10,inner sep=0pt,minimum size=6mm]
 \node[nd] (1) at (-8,0)  {$i_1$};
 \node[nd] (2) at (0,0) {$i_2$};
 \node[nd] (3) at (0,-8) {$i_3$};
 \node[nd] (4) at (-8,-8) {$i_4$};
 \path[every node/.style={font=\sffamily\footnotesize}]
    (1) edge[ right=15, ultra thick] (2)

    (2) edge[ right=15, thick] (3)

    (3) edge[ right=15,  ultra thick] (4)

    ;
 \end{tikzpicture}
    \hspace*{2.5cm}
  \begin{tikzpicture}[-, >=stealth',scale = 0.15, inner sep=2mm]
\tikzstyle{nd}=[circle,draw,fill=black!10,inner sep=0pt,minimum size=6mm]
 \node[nd] (1) at (-8,0)  {$j_1$};
 \node[nd] (2) at (0,0) {$j_2$};
 \node[nd] (3) at (-8,-8) {$j_3$};
 \path[every node/.style={font=\sffamily\footnotesize}]
    (1) edge[right=15, ultra thick] (2)

    (1) edge[right=15, thick] (3)
    ;
 \end{tikzpicture}
\vspace*{-0.4cm}
\end{tabular}
\end{center}
\caption{\label{fig:scheme}
An example of the first two rounds of an IKEP with $N=\{1,2,3\}$. Round~1 is displayed on the left, and round~2 on the right. In this example, both rounds are the same, irrespectively of the solution concept that we use for the initial allocations. This is because round~1 has a unique maximum matching.}\label{f-2sec}
\end{figure}
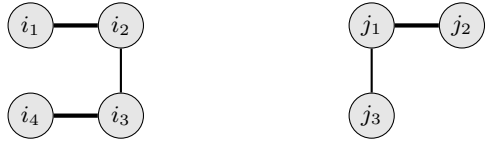

\medskip
\noindent
{\it Example.} 
In Fig.~\ref{f-2sec}, compatibility graphs for two rounds of an IKEP consisting of three countries is displayed, so $N=\{1,2,3\}$.

First assume that we use the Shapley value for the initial allocations.
Let $V_1=\{i_1\}$, $V_2=\{i_2, i_3\}$ and $V_3=\{i_4\}$ in round~$1$. Note that ${\cal M}^1=\{M^1\}$ with $M^1=\{(i_1,i_2),(i_3,i_4)\}$. So, in round~$1$, we need to use $M^1$, as $M^1$ is the only maximum matching in this round. Recall that $c^1=(0,0,0)$. Then $x^1=y^1=(\frac{2}{3}, \frac{8}{3}, \frac{2}{3})$. Moreover, $s(M^1)=(1,2,1)$ and $c^2=y^1-s(M^1)=(-\frac{1}{3}, \frac{2}{3}, -\frac{1}{3})$.
Hence, after round~$1$, all patient-donor pairs $i_1,\ldots, i_4$ have been helped and leave the IKEP. 
Let $V_1=\{j_1\}$, $V_2=\{j_2\}$ and $V_3=\{j_3\}$ in round~$2$. Note that ${\cal M}^2=\{M^2,M^2_*\}$ with $M^2=\{j_1j_2\}$ and $M^2_*=\{j_1j_3\}$. So, in round~$2$, we must choose between using $M^2$ or $M^2_*$, and this choice will be determined by which maximum matching will be closer to the target allocation $x^2$. Note that
$y^2=(\frac{4}{3}, \frac{1}{3}, \frac{1}{3})$. Hence, $x^2=y^2+c^2=(1,1,0)$ and we must choose $M^2=\{(j_1,j_2)\}$, which has $s(M^2)=(1,1,0)$. Consequently, $c^3=x^2-s(M^2)=(0,0,0)$.

Using, for example, the nucleolus for the initial allocations yields the same for round 1 (as we must use $M^1$). However, in round 2, $y^2=(2,0,0)$. Hence, $x^2=y^2+c^2=(\frac{5}{3}, \frac{2}{3}, -\frac{1}{3})$ 
meaning that again we must pick $M^2$. But now, $c^3=x^2-s(M^2)=(\frac{2}{3}, -\frac{1}{3}, -\frac{1}{3})$. This means that in round~3, the nucleolus may lead to choosing a different maximum matching from ${\cal M}^3$. In that case, different patient-donor pairs may leave the IKEP. Consequently, the compatibility graph for round~4 may be different from the compatibility graph for round~4 if the Shapley value had been used. 

We now return to our remark regarding the possible accumulation of credits, as for every country $p\in N$ and round $h\geq 2$, it holds that $c^h_p=\sum_{t=1}^{h-1}(y_p^t - s_p(M^t))$. Suppose that round~$3$ and every future round looks exactly the same as round~2, and suppose that the nucleolus is used as the target allocation. Then, $c^h_{i_1}=c^{h-1}_{i_1}+1$ for $h\geq 3$.  This is clearly not desirable. We monitor whether this situation happens in our simulations. However, as mentioned, we did not see this kind of behaviour occur (see Section~\ref{s-ca}).
\dia

\medskip
\noindent
{\bf Alternative Credit System.} A new approach, which preserves superadditivity and convexity, is to define for some round~$h$ in an IKEP the {\it credit-adjusted game} $\overline{v}^h$ with $$\overline{v}^h(S) = v^h(S) + \sum_{p \in S}c^h_p$$ for every $S\subseteq N$.
For solution concepts that are {\it covariant (under strategic equivalence)}, i.e., that prescribe the same set of allocations 
both for $(N,v)$ and for $(N,\gamma v + \delta)$ for every $\gamma>0$ and every $\delta\in \R^n$, this credit based system works in exactly the same way as before. All solution concepts that we consider have this property except for one:  the (normalized) Banzhaf value. Therefore, in order to research whether this alternative way of incorporating credits improves the stability of an IKEP, we only have to perform an extra set of simulations for the (normalized) Banzhaf value.

\section{Computing a Lexicographically Minimal Maximum Matching}\label{s-poly}

Let $(N,v)$ be a partitioned matching game with a set $V$ of patient-donor pairs.
Let ${\cal M}$ be the set of maximum matchings in the corresponding compatibility graph $D$, and let $x$ be an allocation.
In this section we will give our algorithm {\tt Lex-Min} that we use for computing a maximum matching from ${\cal M}$ that is lexicographically minimal for $x$. 
This algorithm computes for a partitioned matching game $(N,v)$ and allocation $x$, strictly decreasing values $d_1,\ldots, d_t$ for some integer $t\geq 1$, and returns a matching $M\in {\cal M}$ that is lexicographically minimal for~$x$. For computing $d_1,\ldots,d_t$, the algorithm calls the polynomial-time algorithm provided by the following lemma from~\cite{BBKP22} (see also~\cite{BBKPP}).

\begin{lemma}[\cite{BBKP22}]\label{l-interval}
Given a partitioned matching game $(N,v)$ on a graph $G=(V,E)$ with a positive edge weighting $w$ and and partition ${\cal V}$ of $V$,
and intervals $I_1, \dots, I_n$, it is possible in $O(|V|^3)$-time to decide if there exists a matching $M\in {\mathcal M}$ with $s_p(M)\in I_p$ for $p=1,\ldots,n$, and to find such a matching (if it exists).
\end{lemma}

\noindent \rule{\textwidth}{0.1mm}\\
\noindent
{\tt Lex-Min}\\[5pt]
\begin{tabular}[h]{lcl}
{\it input} &:& a partitioned matching game $(N,v)$ and an allocation $x$\\
{\it output} &:& a matching $M\in {\cal M}$ that is lexicographically minimal for $x$.
\end{tabular}

\smallskip
\noindent
{\bf Step~1.} Compute the smallest number $d_1 \ge 0$ such that there exists a matching $M \in \mathcal{M}$ with
$|x_p-s_p(M)|  \le d_1 \text{~~for all~~} p \in N$.

\smallskip
\noindent
{\bf Step~2.} Compute a minimal set $N_1 \subseteq N$ (with respect to set inclusion) such that there exists a matching $M \in \mathcal{M}$ with\\[-15pt]
\begin{align}\nonumber
&|x_p-s_p(M)|
= d_1& \text{~~for all~~} p \in N_1\\
&|x_p-s_p(M)|  < d_1& \text{~~for all~~} p \in N\setminus N_1.\nonumber
\end{align}

\smallskip
\noindent
{\bf Step 3.} Proceed in a similar way for $t\geq 1$:
\begin{itemize}
 \item {\bf while} $N_1 \cup \dots \cup N_t \neq N$ {\bf do}
 \begin{itemize}
 \item $t \leftarrow t+1$.
 \item $d_t \leftarrow$ smallest $d$ such that there exists a matching $M \in \mathcal{M}$ with
 \begin{align}\nonumber
 &|x_p-s_p(M)| =
  d_j &\text{~~for all~~} p \in N_j, ~j\le t-1\\
 &|x_p-s_p(M)|  \le d_t &\text{~~for all~~} p \in N\setminus (N_1 \cup \dots \cup\nonumber N_{t-1}).
 \end{align}
 \item $N_t \leftarrow$ inclusion minimal subset of $N\setminus (N_1 \cup \dots \cup N_{t-1})$ such that  there exists a matching $M \in \mathcal{M}$ with
 \begin{align} \nonumber
 &|x_p-s_p(M)|
 =
  d_j &\text{~~for all~~} p \in N_j, ~j\le t-1\\ \nonumber
 &|x_p-s_p(M)|
=
 d_t &\text{~~for all~~} p \in N_t\\ \nonumber
 &|x_p-s_p(M)|  < d_t &\text{~~for all~~} p \in N\setminus (N_1 \cup \dots \cup N_{t}). \nonumber
 \end{align}
 \end{itemize}
\end{itemize}
\smallskip
\noindent
{\bf Step 4.} Return a matching $M\in {\cal M}$ with
$|x_p-s_p(M)|=d_j$ for all $p\in N_j$ and all $j\in \{1,\ldots,t\}$.

\noindent \rule{\textwidth}{0.1mm}\\[5pt]
\noindent
We say that the countries in a set $N\setminus (N_1\cup \cdots \cup N_{t-1})$ are {\it unfinished} and that a country is {\it finished} when it is placed in some $N_t$. Note that {\tt Lex-Min} terminates as soon as all countries are finished. For a correctness proof and running time analysis of {\t Lex-Min} we refer to~\cite{BBKP22} (see also~\cite{BBKPP}).

\begin{theorem}[\cite{BBKP22}]\label{t-lexmin}
The {\tt Lex-Min} algorithm is correct and runs in $O(n|V|^3\log|V|)$ time for a partitioned matching game $(N,v)$ with an allocation $x$.
\end{theorem}

\noindent
Note that {\tt Lex-Min} can also be used for computing a maximum matching $M$ that minimizes the maximum deviation $d_1$ from an allocation~$x$ and as such does not need to be lexicographically minimal.

\section{Simulations Details}\label{s-sim}

In this section we describe our simulations in detail.
Our goals are 
\begin{itemize}
\item [1.] to examine the benefits of using {\tt Lex-Min} instead of taking a maximum matching that minimizes the largest country deviation~$d_1$ from the target allocation $x$
or just an arbitrary matching,
\item [2.] to test the effect of several (sophisticated) solution concepts.
\end{itemize}
We therefore perform simulations for a large number of countries, as we explain below. Moreover, we do this in the follow two settings where:
\begin{itemize}
\item [(i)] countries all have the same size, and
\item [(ii)] countries have three different sizes with ratio small:medium:large$=$1:2:3.
\end{itemize}

\noindent
{\bf Simulation instances.} We first consider setting (i) which is when all countries have the same size.
Using the generator~\cite{Pa21}\footnote{This is an updated version of~\cite{Tr16}, a data generator which is commonly used in experimental studies (see e.g.~\cite{BDNPSS20,DGGKMP20,KPV16,MBGPV20}) but nowadays (e.g. in~\cite{Delorme_etal2021}) it is replaced by~\cite{Pa21}.}
we obtain
100 compatibility graphs $D_1,\ldots,D_{100}$, each with roughly~2000 vertices.\footnote{We set $D_1,\ldots,D_{100}$ to have size~2000 to be able to go up to $n=15$. Initially, we did a simulation with a size of~$1000$ for up to $n=10$ and found similar results. Hence, we believe our choice of 2000 is robust.}

For every $i\in \{1,\ldots,100\}$ we do as follows. For every $n\in \{4,\ldots,15\}$, we perform simulations for $n$ countries. We first partition $V(D_i)$ into $n$ arbitrary sets $V_{i,1},\ldots, V_{i,n}$ that are all of the same size ${2000}/{n}$ (subject to rounding), so $V_{i,p}$ corresponds to the set of patient-donor pairs of country~$p$.

For round~1, we construct a compatibility graph~$D_i^1$ as a subgraph of $D_i$ of size roughly~500. 
So, a quarter of the patient-donor pairs will enter the program in round~1. The remaining patient-donor pairs of $D_i$ will be added
as vertices to the compatibility graph randomly, by a uniform distribution between the remaining rounds. Starting with $D_i^1(n)$ we run a 6-year IKEP with quarterly matching rounds, that is, a simulation that consists of 24 rounds in total. In this way we obtain 24 compatibility graphs $D_i^1(n),\ldots, D_i^{24}(n)$.

Let $M_i^j(n)$ be the maximum matching that we compute for $D_i^j(n)$.
If $j\leq 23$, then we construct $D_i^{j+1}(n)$ as follows.
First, we remove the vertices that are matched by $M_i^j(n)$; the corresponding patient-donor pairs have been helped.
If $j\geq 4$, then we also remove those vertices from $D_i^j(n)$ that are not in $M_i^j(n)$ but that do belong to $D_i^{j-3}(n)$. This is 
because in real life, patients may seek for alternative treatment or may have deceased after being in the KEP for a year.
Finally, we add the vertices that correspond to the patient-donor pairs that were assigned, in advance of the simulation, to enter the program in round~$j+1$.

A {\it (24-round) simulation instance} consists of the data needed to generate a graph $D_i^1(n)$ and its successors $D_i^2(n),\ldots,D_i^{24}(n)$, together with specifications for the initial allocation $y$ and maximum matching~$M$ to be used in each round. Our code for obtaining the simulation instances is available in GitHub repository~\cite{B21}, along with the data from~\cite{Pa21} describing the compatibility graphs and the seeds used for the randomization.

For setting (ii), where country sizes are varying, we do exactly the same except that we impose different restrictions on the sizes of the countries. Namely, for the sets $V_{i,1},\ldots, V_{i,n}$ it holds that approximately $n/3$ are small, that is, have size roughly $2000/8n$ (subject to rounding); approximately $n/3$ are medium, that is, have size roughly $2000/4n$ (twice as large as small countries); and approximately $n/3$ are large (same amount as small), that is, have size roughly $3000/4n$ (three times as large as small).

We now discuss how we computed the initial allocations and maximum matchings.

\medskip
\noindent
{\bf Initial allocations.}
Recall from Section~\ref{s-model} that for the initial allocations~$y$ we use the Shapley value, nucleolus, Banzhaf value, tau value, the benefit value and the contribution value. Recall also from Section~\ref{s-model} that when $y$ is the tau value, we will replace the tau value by the benefit value if the corresponding partitioned matching game is not quasibalanced. Hence, strictly speaking we use a 
hybrid tau value, but we only need to make this replacement in 0.04\% of the games overall; 
see Table~\ref{table:tau*} in Section~\ref{s-convex}.

Let $(N,v)$ be a partitioned matching game defined on some compatibility graph $D_i^j(n)$ with $n$ countries.
For a coalition of countries $S\subseteq N$, $v(S)$ is the size of a maximum matching in the subgraph of $D_i^j(n)$ induced by the vertices of the countries of $S$.
We compute the size of a maximum matching in such a subgraph efficiently, using the package of~\cite{DJK11}.
The contribution value and benefit value can now be efficiently computed, using their definitions.

For the Shapley value and Banzhaf value, we were still able to implement
a naive (brute force) approach relying directly on \eqref{eq-Shapley} (see also Table~\ref{table2} of Section~\ref{s-compreq}). 
For the tau value, we first need to compute the vectors $a$ and $b$. Only $b$ can be computed efficiently, but for computing $a$ we were also still able to use a naive approach that relies directly on its definition. 
However, a naive approach for computing the nucleolus of a partitioned matching games is not possible for the high number of countries we consider. We therefore use
 the \emph{Lexicographical Descent} method of \cite{BFN20}, which is the state-of-the-art method in nucleolus computation.\footnote{Lexicographical descent breaks down the computation of the nucleolus into $O(n)$ linear programs (LPs), which still have exponential size, but can be easier handled, through the solution of small dual relaxations combined with easily generated primal feasible starting points. The run time is still exponential, but in this way we are able to deal with significantly larger problem instances than in previous approaches.}

\smallskip
\noindent
{\bf Solutions.}
As mentioned, we aim to examine the benefits of using solutions maximum matchings prescribed by {\tt Lex-Min} over arbitrary matchings or maximum matchings that minimize the maximum deviation $d_1$ from the target allocation. For computing an arbitrary maximum matching we use one given to us by the package of~\cite{DJK11}. For computing a maximum matching that minimizes only $d_1$ it suffices to perform only the first step of {\tt Lex-Min}.

We implemented {\tt Lex-Min} as follows 
(for the exact computer code and an explanatory pseudocode, see our GitHub repository~\cite{B21}).
Instead of a binary search for finding a deviation~$d_t$, we performed a greedy search for simplicity. We gradually try to decrease the deviations of the countries in a greedy way, starting with the ones that have the largest deviation. We maintain a set of countries already finished, which initially is the empty set. In every iteration we take one of the unfinished countries, say
country~$p$, with the largest deviation $\delta_p=|x_p-s_p(M)|$, where $M\in {\cal M}$ is the last maximum matching we computed. We then try to decrease this deviation without allowing a deviation of another unfinished country to increase above $\delta_p$. If no decrease is possible, then we fix country~$p$ and move to the next unfinished country with the largest deviation. If a decrease is possible, then we consider the newly found maximum matching and repeat the process. That is, we consider again a country (possibly country~$p$) with the largest deviation.

We briefly discuss the running time of the greedy variant of {\tt Lex-min}.
In each iteration, either we finish a country, or we shrink the deviation interval of a country by an integer step. Therefore, the number of iterations is upper bounded by the number $n$ of countries plus the sum of the upper integer part of the initial deviations, which is $O(|V|)$. As an application of Lemma~\ref{l-interval} takes $O(|V|^3)$ time,
 the total running time is $O((|V|+n)|V|^3)$. This running time can be worse than the running time of the binary search version, which is $O(n|V|^3\log|V|)$ by Theorem~\ref{t-lexmin}, as
 $n\log(|V|)$ may be smaller than $|V|+n$. 
 However, for the instances in our simulations the difference in running time appeared to be negligible.

Finally, as the {\tt Lex-Min} algorithm uses Lemma~\ref{l-interval} as a subroutine, we needed to implement the polynomial-time algorithm provided by Lemma~\ref{l-interval} as well, see~\cite{BBKP22} for a brief description of this algorithm, or~\cite{BBKPP} for a full description. As explained in~\cite{BBKP22} (see also~\cite{BBKPP}), applying Lemma~\ref{l-interval} requires solving a maximum weight perfect matching problem. In order to do the latter efficiently, we again used the package of \cite{DJK11}.

\medskip
\noindent
{\bf Credit system.} We aim to
distinguish between the effect of {\tt Lex-min} and the effect of $c$,
and also to distinguish between the effect of {\tt Lex-min} over choosing a maximum matching that minimizes $d_1$ or an arbitrary matching (our baseline).
Note that in the latter case, using the credit function $c$ is meaningless, as in each round we pick an arbitrary maximum matching, independently from what happened in the previous round.
Hence, we run the same simulations for the following five {\it scenarios}, where $y$ is prescribed by one of the following six solution concepts: the Shapley value, nucleolus, Banzhaf value, tau value, benefit value and contribution value.

\begin{enumerate}
\item \emph{arbitary}: $M$ is an arbitrary maximum matching (one computed by the package of~\cite{DJK11})
\item  \emph{d1}: $M$ is a maximum matching that minimizes $d_1$ and $x=y$.
\item \emph{d1+c}: $M$ is a maximum matching that minimizes $d_1$ and $x=y+c$.
\item  \emph{lexmin}: $M$ is the maximum matching returned by {\tt Lex-Min} and $x=y$.
\item \emph{lexmin+c}: $M$ is the maximum matching returned by {\tt Lex-Min} and $x=y+c$.
\end{enumerate}

\noindent
Additionally, we evaluate the effect of taking credit-adjusted games instead of setting $x=y+c$. As explained in Section~\ref{s-model}, this effect can only be measured by taking the Banzhaf value as the initial allocation $\overline{y}$ for the credit-adjusted games (for all the other solution concepts, we obtain the same outcomes). Hence, this leads to two more scenarios:

\begin{enumerate}
\item [6.] \emph{$\overline{d1}$}: $M$ is a maximum matching that minimizes $d_1$ and $x=\overline{y}$.
\item [7.] \emph{$\overline{\mbox{lexmin}}$}: $M$ is the maximum matching returned by {\tt Lex-Min} and $x=\overline{y}$.
\end{enumerate}
Hence, in total, we run the same set of simulations for 25+2=27 different combinations of scenarios and initial allocations.

\medskip
\noindent
{\bf Computational environment and scale.}
We ran all simulations on a desktop PC with AMD Ryzen 9 5950X 3.4 GHz CPU and 128~GB of RAM, running on Windows 10 OS and C++ implementation in Visual Studio. Our code~\cite{B21}
use the open-source code~\cite{B18} of the Lexicographical Descent method for computing the nucleolus.
The scale of our experiments for IKEPs is {\it unprecedented}: our total number of 24-round simulation instances 
is equal to $2\times 27  \times 12 \times 100 = 64800$, (namely, two different settings (same or varying country sizes, 27 combinations of scenarios and initial allocations,  twelve country sizes~$n$;  and 100 initial compatibility graphs $D_i$).

\medskip
\noindent
{\bf Evaluation measures.}
To measure balancedness we do as follows. After the 24 matching runs of a single instance, we will have a total target allocation $x^*$, which is defined as the sum of the 24 target allocations, and a maximum matching $M^*$, which is the union of the chosen matchings in each of the 24 matching runs. Note that the total number of kidney transplants is $2|M^*|$.
We now define the {\it total relative deviation} as
$$\frac{\sum_{p \in N} |x_p^* - s_p(M^*)|}{2|M^*|}.$$
Recall that for each triple that consists of a country set size, choice of initial allocation and choice of scenario,
 we run 100 instances. We take the average of the 100 relative total deviations. This gives us the \emph{average total relative deviation}.
 
Apart from using the average total relative deviation as our evaluation measure, we also took the {\it maximum relative deviation}, which is defined as 
$$\frac{\max_{p \in N} |x_p^* - s_p(M^*)|}{2|M^*|},$$ 
leading to the {\it average maximum relative deviation} as our second evaluation measure.

\section{Main Results}\label{s-results}

\begin{figure}
		\resizebox{\textwidth}{!}{
			\includegraphics{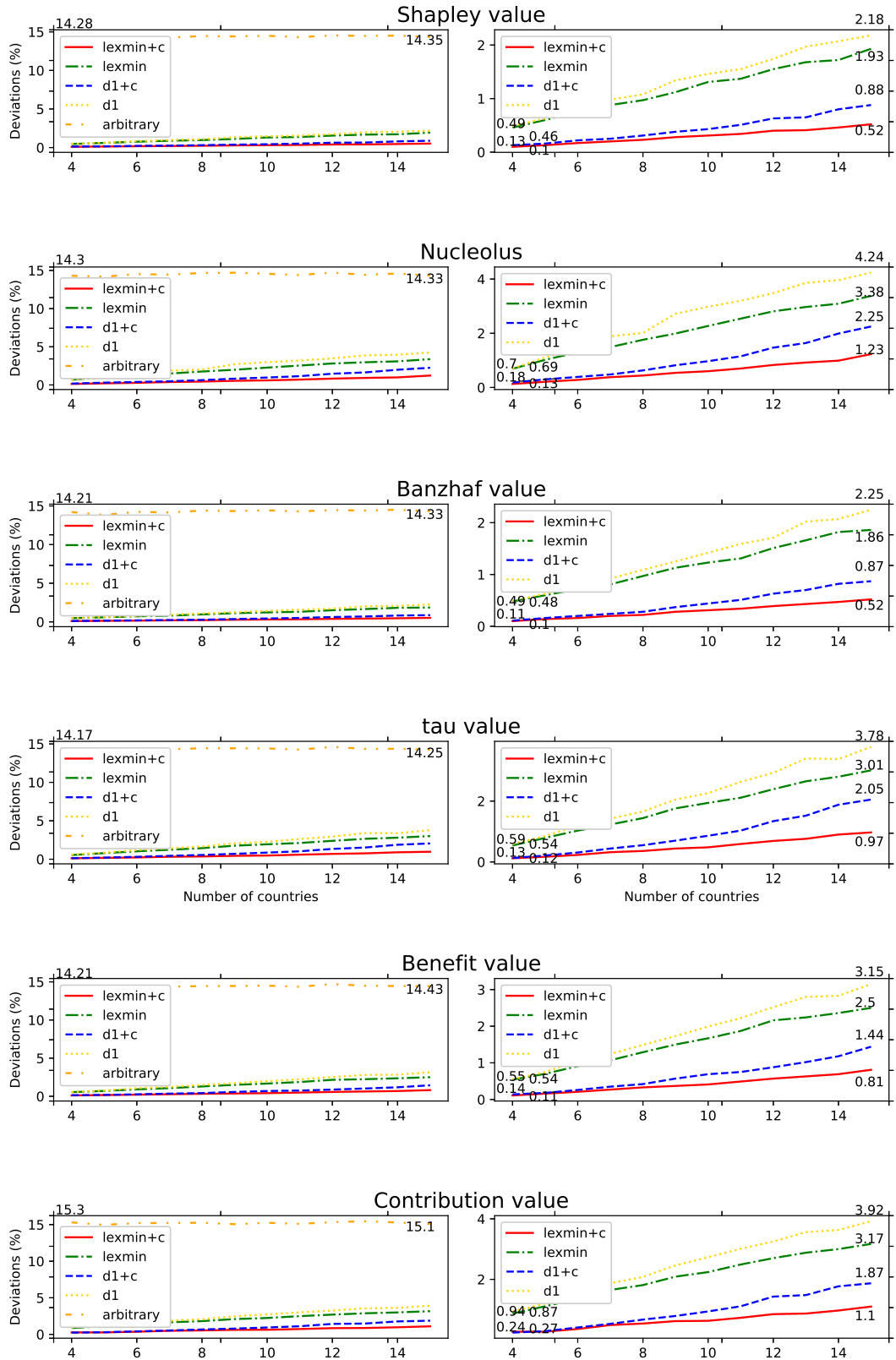}}
			\caption{Average total relative deviations for the situation where all countries have the same size. The number of countries $n$ is ranging from $4$ to $15$. The figures on the right side zoom in on the figures from the left side by removing the results for the arbitrary matching scenario.}\label{fig1}
\vspace*{-0.6cm}
\end{figure}

In Figure~\ref{fig1} we display the main results for the situation where all countries have the same size and when we use the average total relative deviation as our evaluation measure. As expected, using an arbitrary maximum matching in each round makes the kidney exchange scheme significantly more unbalanced, with average total relative deviations above 13.8\% for all initial allocations~$y$. 

From Figure~\ref{fig1} we can compute the {\it relative improvement} of \emph{lexmin+c} over \emph{d1+c}. For example, for $n=15$, this percentage is
$(2.05-0.97)/2.05=${\bf 52.49}\% for the tau value, whereas for the other solution concepts it is 45.5\% (nucleolus); 44\% (benefit value); 41\% (contribution value); 40\% (Shapley value); and 40\% (Banzhaf value).
Considering the average improvement over $n=4,\ldots,15$ yields percentages of 37\% (tau value); 36\% (nucleolus); 31\% (benefit value); 30\% (Shapley value);  27\% (Banzhaf value); and 24\% (contribution value).

From Figure~\ref{fig1} we can also compare \emph{lexmin+c} with \emph{lexmin}, and \emph{d1+c} with \emph{d1}. We see that using~$c$ has a substantial effect.
Whilst \emph{lexmin} ensures that allocations stay close to the target allocations, the role of $c$ is to keep the deviations small and to guarantee fairness for the participating countries over a long time period.

\begin{figure}
		\resizebox{\textwidth}{!}{
			\includegraphics{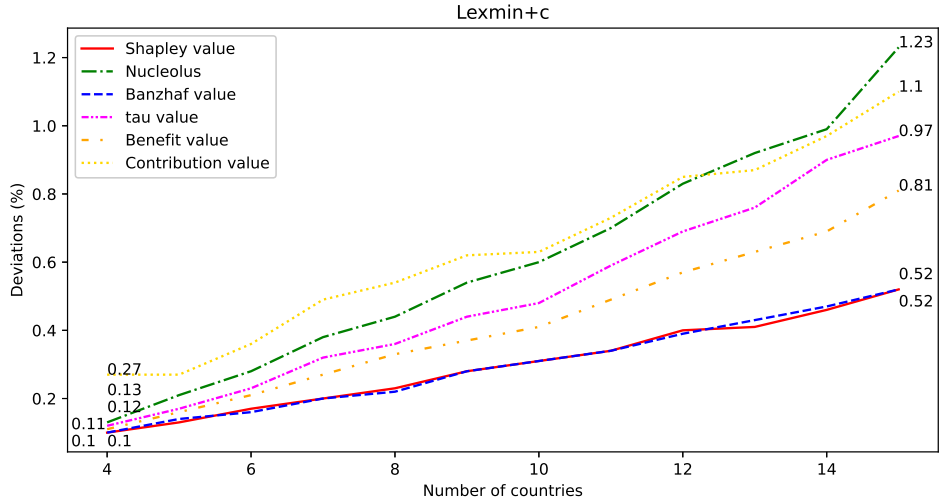}}
			\caption{Displaying the six \emph{lexmin+c} graphs of Figure~\ref{fig1} in one plot.}\label{fig4}
\vspace*{-0.38cm}
\end{figure}

Our main conclusion from Figure~\ref{fig1} is that using \emph{lexmin+c} yields the lowest average total relative deviation for all six initial allocations~$y$, with larger differences when the number of countries is growing.
In Figure~\ref{fig4} we displayed the six \emph{lexmin+c} graphs of Figure~\ref{fig1} in one plot in order to compare them with each other.
As mentioned, the choice for initial allocation is up to the policy makers of the IKEP.
However, from Figure~\ref{fig4} we see that the Shapley value and the Banzhaf value in the \emph{lexmin+c} scenario consistently provides the smallest deviations from the target allocations ({\bf 0.52\%} for $n=15$), while the contribution value for $n \leq 12$ and the nucleolus for $n\geq 13$ perform the worst. The latter result is perhaps somewhat surprising given the sophisticated nature of the nucleolus.

\begin{figure}[!h]
		\resizebox{\textwidth}{!}{
			\includegraphics{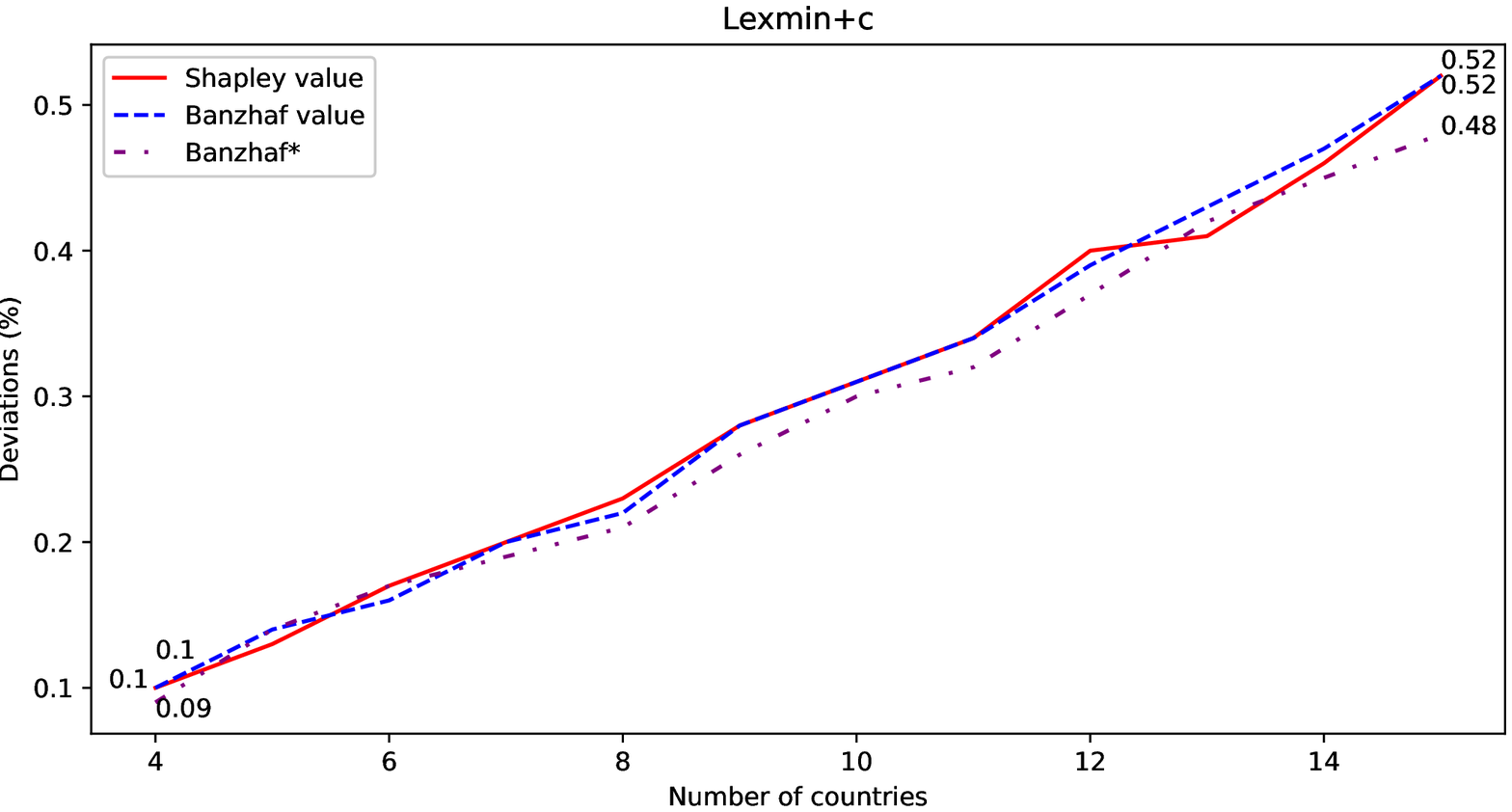}}
			\caption{Comparing the \emph{lexmin+c} graphs for the Shapley value and Banzhaf value from Figure~\ref{fig1} with the one for the Banzhaf* value (same country sizes).}\label{fig4b}
\vspace*{-0.38cm}
\end{figure}

\begin{figure}[!h]
		\resizebox{\textwidth}{!}{
			\includegraphics{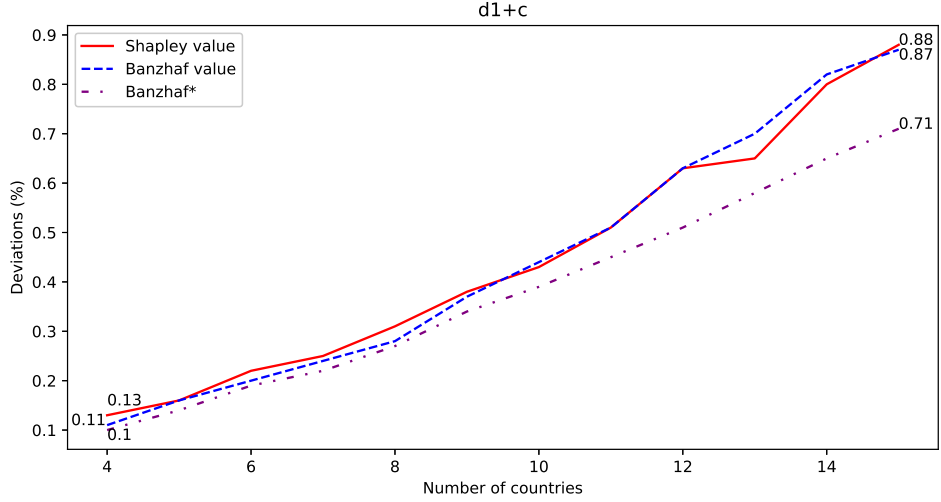}}
			\caption{Comparing the \emph{d1+c} graphs for the Shapley value and Banzhaf value from Figure~\ref{fig1} with the one for the Banzhaf* value (same country sizes).}\label{fig4c}
\vspace*{-0.38cm}
\end{figure}

\medskip
\noindent
We now turn to the Banzhaf value for the credit-adjusted games, which we denote as the Banzhaf* value. Recall that choosing any of the other solution concepts as initial allocation will lead to the same results as for the original games, an only for the Banzhaf value the two different credit systems may give different results. Figure~\ref{fig4b} show that the latter is indeed the case. It displays the \emph{lexmin+c} graphs for the Shapley value and (original) Banzhaf value from Figure~\ref{fig1} and compares them with the \emph{lexmin+c} graph for the Banzhaf* value. Figure~\ref{fig4c} does the same for the three \emph{d1+c} graphs. Both figures show that the Banzhaf* value behaves better than the Shapley value and Banzhaf value; however the differences are very small (at most 0.04\% for lexmin+c and 0.19\% for d1+c).

If we use our second evaluation measure,  the average maximum relative deviation, then we obtain similar results and can draw the same conclusions; we refer to Appendix~\ref{a-max} for the corresponding figures.

\begin{figure}
		\resizebox{\textwidth}{!}{
			\includegraphics{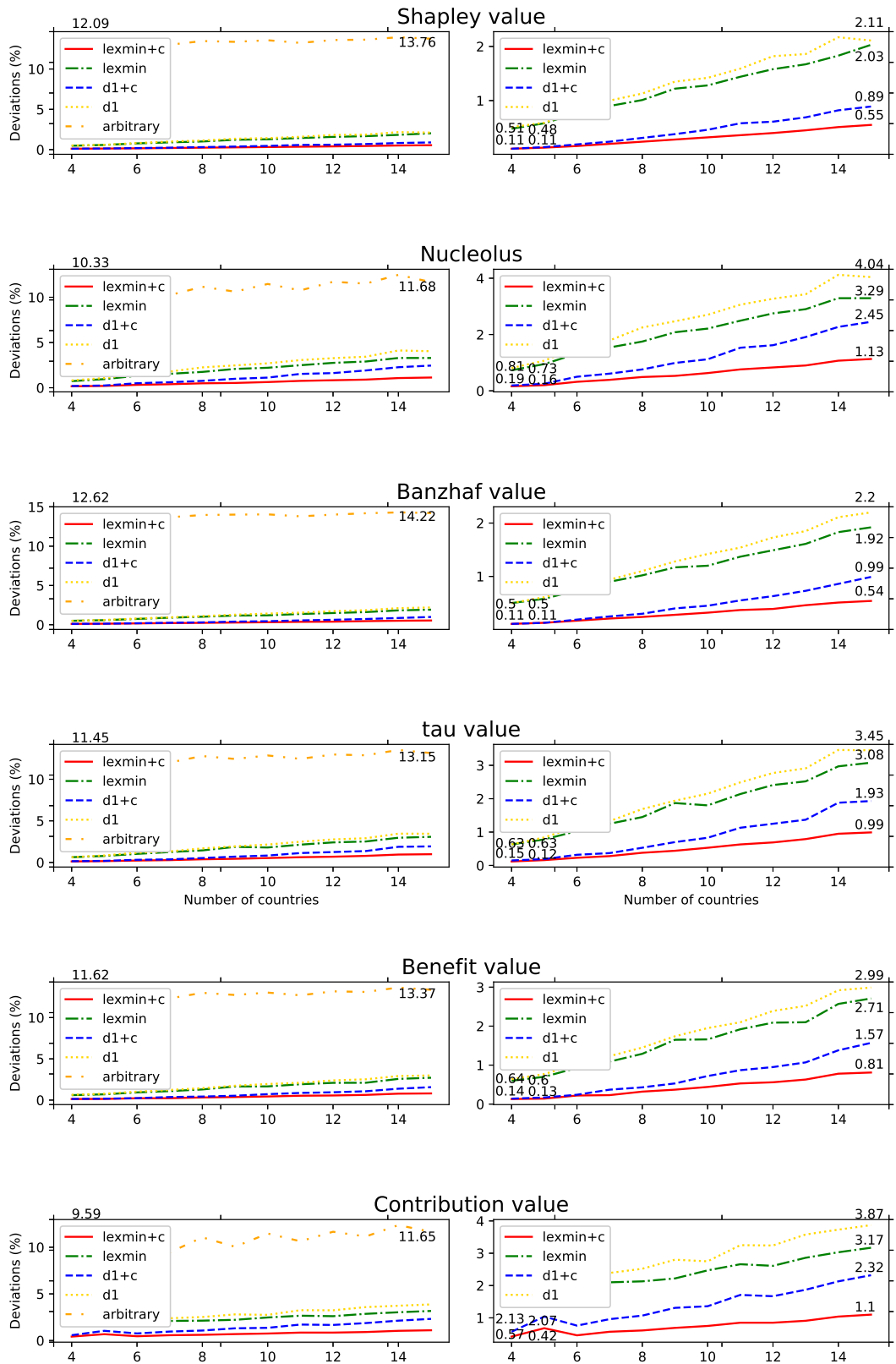}}
			\caption{Average total relative deviations for the situation where the countries vary in size. The number of countries $n$ is ranging from $4$ to $15$. The figures on the right side zoom in on the figures from the left side by removing the results for the arbitrary matching scenario.}\label{fig1var}
\end{figure}

\begin{figure}
		\resizebox{\textwidth}{!}{
			\includegraphics{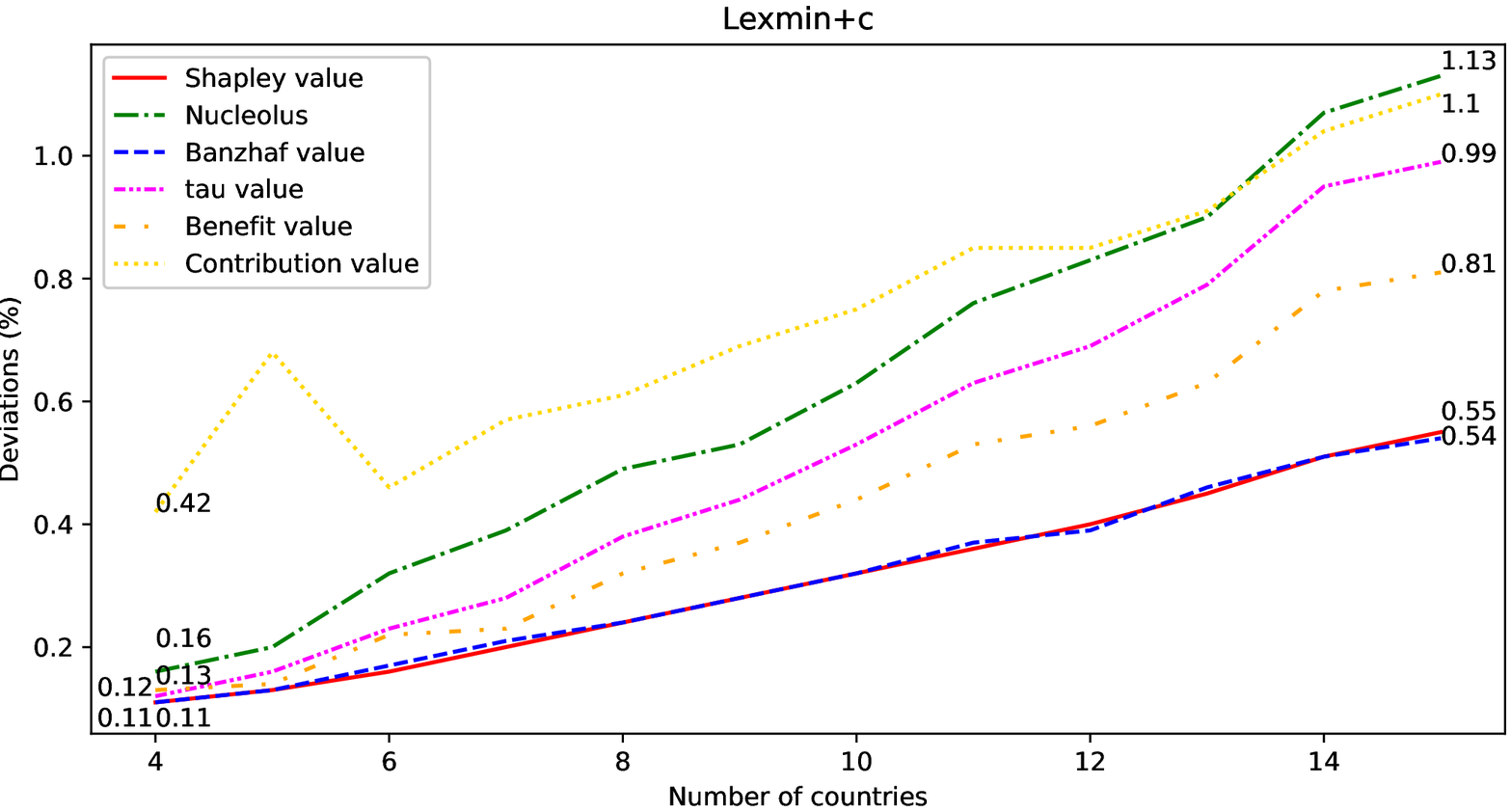}}
			\caption{Displaying the six \emph{lexmin+c} graphs of Figure~\ref{fig1var} in one plot.}\label{fig4var}
\end{figure}

\medskip
\noindent
We now turn to the situation of varying country sizes and perform the same simulations as before. We can draw exactly the same conclusions with different percentages. That is,
from Figure~\ref{fig1var} we see that using an arbitrary maximum matching in each round makes the kidney exchange scheme significantly more unbalanced, with average total relative deviations above 8.15\% for all initial allocations~$y$. Moreover,
from Figure~\ref{fig1var} we can also compute the {\it relative improvement} of \emph{lexmin+c} over \emph{d1+c}. For example, for $n=15$, this percentage is
$(1.13-2.45)/2.45=${\bf 54}\% for the nucleolus, whereas for the other solution concepts it is 53\% (contribution value); 49\% (tau value); 48\% (benefit value); 46\% (Banzhaf value); and 38\% (Shapley value).
Considering the average improvement over $n=4,\ldots,15$ now yields percentages of 44\% (contribution value); 41\% (nucleolus); 35\% (tau value); 32\% (benefit value); 25\% (Shapley value); and 25\% (Banzhaf value).
Compare \emph{lexmin+c} with \emph{lexmin}, and \emph{d1+c} with \emph{d1}, we see again that using~$c$ has a substantial effect.
Our main conclusion from Figure~\ref{fig1var} is again that using \emph{lexmin+c} yields the lowest average total relative deviation for all six initial allocations~$y$, with larger differences when the number of countries is growing. However, from Figure~\ref{fig4var} we see again that the Shapley value and the Banzhaf value in the \emph{lexmin+c} scenario consistently provides the smallest deviations from the target allocations ({\bf 0.55\%} and {\bf 0.54\%} for $n=15$), while the contribution value for $n \leq 13$ and the nucleolus for $n\geq 14$ perform the worst. 

\begin{figure}[!h]
		\resizebox{\textwidth}{!}{
			\includegraphics{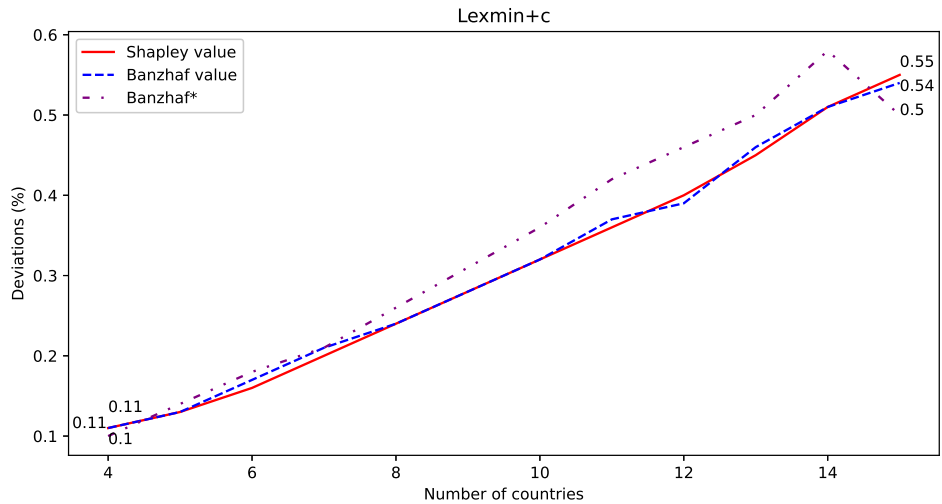}}
			\caption{Comparing the \emph{lexmin+c} graphs for the Shapley value and Banzhaf value from Figure~\ref{fig1} with the one for the Banzhaf* value (varying country sizes).}\label{fig4bvar}
\vspace*{-0.38cm}
\end{figure}

\begin{figure}[!h]
		\resizebox{\textwidth}{!}{
			\includegraphics{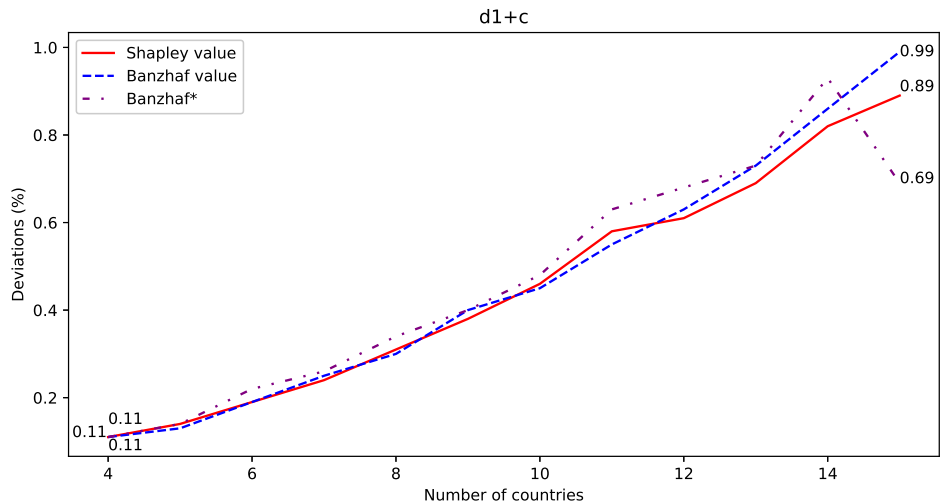}}
			\caption{Comparing the \emph{d1+c} graphs for the Shapley value and Banzhaf value from Figure~\ref{fig1} with the one for the Banzhaf* value (varying country sizes).}\label{fig4cvar}
\vspace*{-0.38cm}
\end{figure}

\medskip
\noindent
Turning now to the credit-adjusted games and the Banzhaf* value, the situation with only the Banzhaf value producing different results among the tested allocations, and only in scenarios \emph{lexmin+c} and \emph{d1+c} (since without credits the games are the same), naturally remains the same. However, as shown in Figures~\ref{fig4bvar} and~\ref{fig4cvar}, the behaviour of Banzhaf* value differs under the varying country sizes, in one part performing slightly worse than the Shapley value and the original Banzhaf value for $n \leq 14$ (the differences are again very small: within 0.08\% and 0.11\% for lexmin+c and d1+c respectively), but outperforming both for $n=15$, by at most 0.05\% for lexmin+c and 0.3\% for d1+c.

If we use the average maximum relative deviation instead of the average total relative deviation, then again we obtain similar results and can draw the same conclusions; see again Appendix~\ref{a-max} for the corresponding figures.

\section{Evaluation of Further Aspects}\label{s-evaluation}

In this section we evaluate some other aspects of our simulations. Given that the results of our simulations for varying countries were similar to the results of our simulations for same country sizes, we only evaluated these aspects for the situation, in which all countries have the same size.

\subsection{No Cooperation}

It is a natural question to what extent cooperation between countries helps. Table~\ref{table:details} shows that cooperation leads to a significantly larger number of total kidney transplants than non-cooperation. This is especially the case when more and more countries are participating in the IKEP. In particular, Table~\ref{table:details} shows that a gain of 2.86 times as many kidney transplants can be obtained in the case where the number of countries is~$15$. Hence, Table~\ref{table:details} provides strong evidence for forming large IKEPs.

We also note the following. Theoretically, a change in scenario may result in a change in maximum matching size (total number of kidney transplants). However, Table~\ref{table:details} shows that these differences turn out to be negligible (between 0.01\% and 0.1\% on average).

\begin{table}
\vspace*{-0.5cm}
\begin{center}
\setlength{\tabcolsep}{3.7pt}
\scriptsize
\begin{tabular}{|r|cccccc|}
\hline
Allocations \& scenarios / Countries & 4 & 5 & 6 & 7 & 8 & 9  \\
\hline
Without cooperation  & 1124.28 & 974.56 & 850.60 & 759.40 & 687.28 & 628.92 \\ \hline
 \textbf{ Shapley value: } &&&&&&\\
lexmin+c  & 108.6\% & 125.3\% & 143.1\% & 159.7\% & 175.8\% & 191.7\% \\
lexmin  & 108.6\% & 125.3\% & 143.1\% & 159.7\% & 175.9\% & 191.6\% \\
d1+c  & 108.7\% & 125.3\% & 143.0\% & 159.6\% & 176.0\% & 191.4\% \\
d1  & 108.7\% & 125.3\% & 143.1\% & 159.7\% & 175.8\% & 191.6\% \\
arbitrary matching  & 108.5\% & 125.2\% & 142.9\% & 159.5\% & 175.7\% & 191.5\% \\ \hline
 \textbf{ Nucleolus: } &&&&&&\\
lexmin+c  & 108.7\% & 125.4\% & 143.1\% & 159.5\% & 175.8\% & 191.6\% \\
lexmin  & 108.6\% & 125.4\% & 143.1\% & 159.6\% & 175.9\% & 191.6\% \\
d1+c  & 108.7\% & 125.4\% & 142.9\% & 159.7\% & 175.9\% & 191.5\% \\
d1  & 108.6\% & 125.4\% & 143.1\% & 159.7\% & 175.9\% & 191.6\% \\
arbitrary matching  & 108.5\% & 125.2\% & 142.9\% & 159.5\% & 175.7\% & 191.5\% \\ \hline
 \textbf{ Banzhaf: } &&&&&&\\
lexmin+c & 108.7\% & 125.3\% & 143.0\% & 159.8\% & 175.8\% & 191.7\% \\
lexmin & 108.6\% & 125.2\% & 143.0\% & 159.7\% & 175.9\% & 191.6\% \\
d1+c & 108.7\% & 125.3\% & 143.0\% & 159.6\% & 175.9\% & 191.9\% \\
d1 & 108.5\% & 125.2\% & 143.0\% & 159.8\% & 175.9\% & 191.7\% \\
arbitrary matching & 108.5\% & 125.2\% & 142.9\% & 159.5\% & 175.7\% & 191.5\% \\
 \hline
 \textbf{ tau: } &&&&&&\\
lexmin+c & 108.7\% & 125.4\% & 142.9\% & 159.7\% & 176.0\% & 191.7\% \\
lexmin & 108.6\% & 125.3\% & 143.1\% & 159.7\% & 175.8\% & 191.6\% \\
d1+c & 108.6\% & 125.4\% & 143.1\% & 159.7\% & 175.9\% & 191.7\% \\
d1 & 108.7\% & 125.4\% & 143.0\% & 159.8\% & 176.0\% & 191.8\% \\
arbitrary matching & 108.5\% & 125.2\% & 142.9\% & 159.5\% & 175.7\% & 191.5\% \\ \hline
 \textbf{ Benefit value: } &&&&&&\\
lexmin+c  & 108.6\% & 125.4\% & 143.0\% & 159.7\% & 176.0\% & 191.7\% \\
lexmin  & 108.6\% & 125.5\% & 142.9\% & 159.7\% & 176.0\% & 191.7\% \\
d1+c  & 108.7\% & 125.4\% & 143.1\% & 159.7\% & 175.9\% & 191.9\% \\
d1  & 108.6\% & 125.3\% & 143.1\% & 159.7\% & 175.9\% & 191.6\% \\
arbitrary matching  & 108.5\% & 125.2\% & 142.9\% & 159.5\% & 175.7\% & 191.5\% \\ \hline 
 \textbf{ Contribution value: } &&&&&&\\
lexmin+c  & 108.6\% & 125.4\% & 143.1\% & 159.6\% & 176.0\% & 191.6\% \\
lexmin  & 108.6\% & 125.4\% & 143.0\% & 159.7\% & 175.9\% & 191.8\% \\
d1+c  & 108.7\% & 125.3\% & 143.1\% & 159.7\% & 175.9\% & 191.6\% \\
d1  & 108.6\% & 125.3\% & 143.0\% & 159.6\% & 176.2\% & 191.8\% \\
arbitrary matching  & 108.5\% & 125.2\% & 142.9\% & 159.5\% & 175.7\% & 191.5\% \\ \hline
 \hline
Allocations \& scenarios / Countries & 10 & 11 & 12 & 13 & 14 & 15 \\ \hline
Without cooperation  & 583.84 & 538.68 & 497.72 & 474.92 & 440.52 & 421.88 \\ \hline
 \textbf{ Shapley value: } &&&&&&\\
lexmin+c  & 209.4\% & 224.0\% & 240.6\% & 253.5\% & 270.4\% & 286.3\% \\
lexmin  & 209.5\% & 224.1\% & 240.4\% & 253.3\% & 270.3\% & 286.0\% \\
d1+c  & 209.3\% & 224.2\% & 240.6\% & 253.1\% & 270.0\% & 286.3\% \\
d1  & 209.2\% & 224.3\% & 240.5\% & 253.2\% & 270.0\% & 286.0\% \\
arbitrary matching  & 209.2\% & 224.0\% & 240.2\% & 253.2\% & 270.1\% & 285.7\% \\ \hline
 \textbf{ Nucleolus: } &&&&&&\\
lexmin+c  & 209.2\% & 224.3\% & 240.6\% & 253.3\% & 270.1\% & 286.2\% \\
lexmin  & 209.3\% & 224.3\% & 240.6\% & 253.3\% & 270.2\% & 286.2\% \\
d1+c  & 209.3\% & 224.2\% & 240.4\% & 253.6\% & 270.4\% & 286.2\% \\
d1  & 209.4\% & 224.2\% & 240.5\% & 253.6\% & 270.3\% & 286.4\% \\
arbitrary matching  & 209.2\% & 224.0\% & 240.2\% & 253.2\% & 270.1\% & 285.7\% \\ \hline
 \textbf{ Banzhaf: } &&&&&&\\
lexmin+c & 209.3\% & 224.5\% & 240.4\% & 253.3\% & 270.1\% & 286.2\% \\
lexmin & 209.1\% & 224.3\% & 240.5\% & 253.0\% & 270.0\% & 285.9\% \\
d1+c & 209.3\% & 224.5\% & 240.6\% & 253.2\% & 270.2\% & 286.3\% \\
d1 & 209.4\% & 224.2\% & 240.5\% & 253.1\% & 270.3\% & 286.1\% \\
arbitrary matching & 209.2\% & 224.0\% & 240.2\% & 253.2\% & 270.1\% & 285.7\% \\ \hline
 \textbf{ tau: } &&&&&&\\
lexmin+c & 209.4\% & 224.2\% & 240.5\% & 253.5\% & 270.5\% & 286.2\% \\
lexmin & 209.7\% & 224.4\% & 240.7\% & 253.5\% & 270.3\% & 286.4\% \\
d1+c & 209.5\% & 224.5\% & 240.7\% & 253.7\% & 270.4\% & 286.3\% \\
d1 & 209.6\% & 224.4\% & 240.7\% & 253.4\% & 270.3\% & 286.2\% \\
arbitrary matching  & 209.2\% & 224.0\% & 240.2\% & 253.2\% & 270.1\% & 285.7\% \\ \hline
 \textbf{ Benefit value: } &&&&&&\\
lexmin+c  & 209.6\% & 224.1\% & 240.6\% & 253.2\% & 270.4\% & 286.2\% \\
lexmin  & 209.4\% & 224.4\% & 240.6\% & 253.7\% & 270.3\% & 286.3\% \\
d1+c  & 209.4\% & 224.3\% & 240.6\% & 253.7\% & 270.1\% & 286.2\% \\
d1  & 209.4\% & 224.4\% & 240.7\% & 253.7\% & 270.3\% & 286.1\% \\
arbitrary matching  & 209.2\% & 224.0\% & 240.2\% & 253.2\% & 270.1\% & 285.7\% \\ \hline
 \textbf{ Contribution value: } &&&&&&\\
lexmin+c  & 209.4\% & 224.2\% & 240.6\% & 253.3\% & 270.3\% & 286.3\% \\
lexmin  & 209.5\% & 224.2\% & 240.5\% & 253.4\% & 270.2\% & 286.2\% \\
d1+c  & 209.6\% & 224.3\% & 240.5\% & 253.4\% & 270.2\% & 286.1\% \\
d1  & 209.4\% & 224.2\% & 240.5\% & 253.6\% & 270.4\% & 286.0\% \\
arbitrary matching  & 209.2\% & 224.0\% & 240.2\% & 253.2\% & 270.1\% & 285.7\% \\ \hline
\end{tabular}
\normalsize
\vspace*{2mm}
\caption{For $n=4,\ldots,15$, the improvement on the average number of kidney transplants if cooperation is allowed. For example, if $n=4$, $y$ is the Shapley value and the scenario is \emph{lexmin+c}, then the average number of kidney transplants changes from $1124.28$ (no cooperation) to 
$1.086 \times 1124.28 = 1220.97$.} 
\label{table:details}
\end{center}
\end{table}

\begin{figure}
		\resizebox{\textwidth}{!}{
			\includegraphics{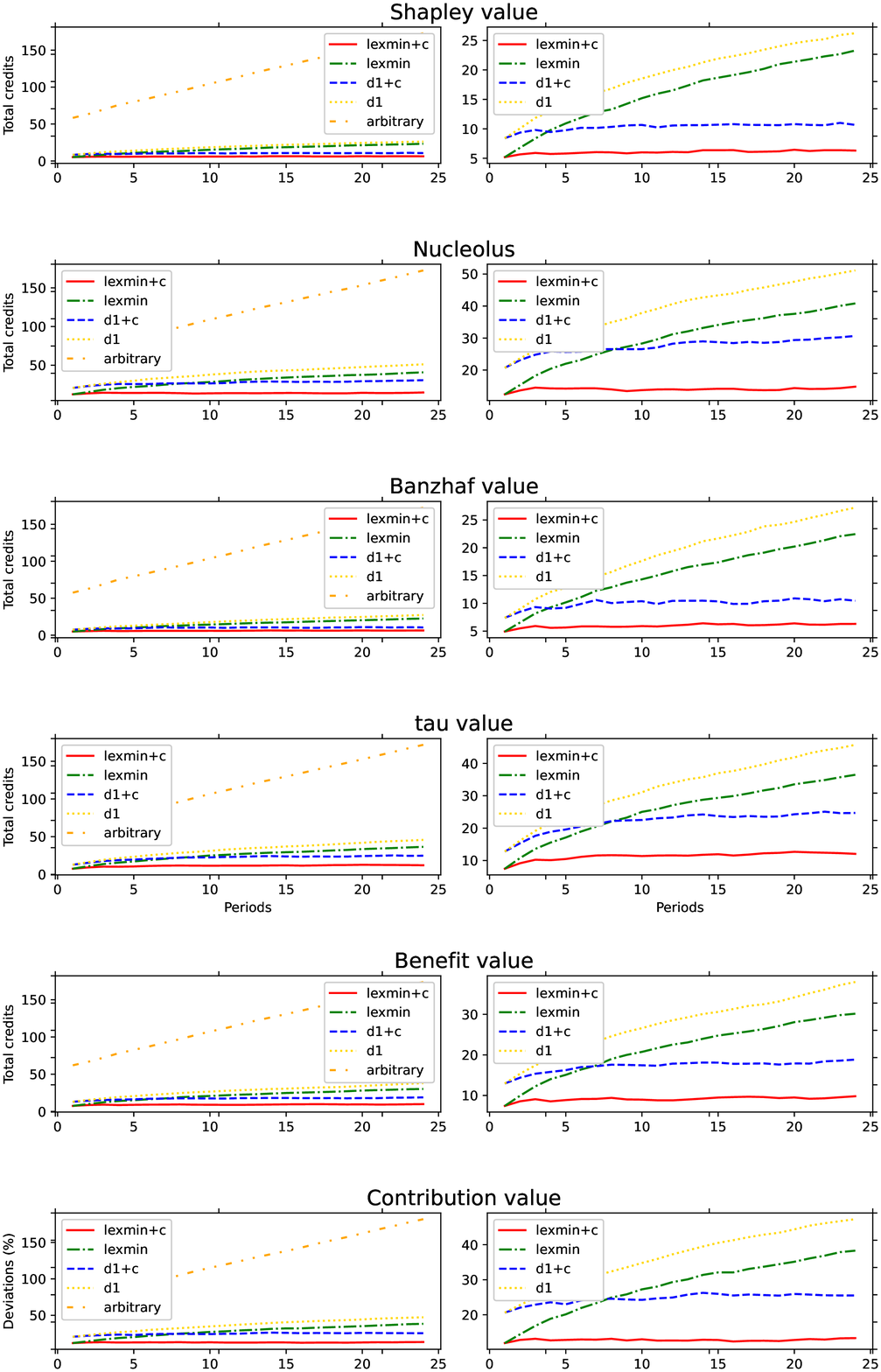}}
			\caption{\label{fig3} The average accumulated deviation over the 24 rounds
when the number of countries $n=15$. The right side of the figure is taken from the left side after omitting the additional scenario where arbitrary maximum matchings are chosen.}
\end{figure}

\subsection{Credit Accumulation}\label{s-ca}

In Section~\ref{s-model}, we gave a theoretical example where credits build up over time for a certain country and are essentially meaningless.
However, this behaviour did not happen in any of our 24-round simulations. 
We performed for every number $n$ of countries with $n\in \{4,\ldots,15\}$,
 a refined analysis, just to verify if such behaviour could be expected if the number of rounds is larger than~$24$. 

First recall that for a single instance, $c^h_p = x^h_p - y^h_p$ and also that $c^h_p=\sum_{t=1}^{h-1}(y_p^t - s_p(M^t))$ for country~$p$ and round~$h\geq 2$. That is, credits are the difference between the initial and target allocations in each round, as well as the accumulation of the deviations from the initial allocations. The latter is the accumulation we would like to avoid from happening by using the credit function~$c$. In order to assess credit accumulation over time, we define the {\it average accumulated deviation} at round~$h$ as
the average of ${\sum_{p \in N} |c_p^h|}$ 
over the 100 instances corresponding to a certain scenario and choice of initial allocation.

In Figure~\ref{fig3} we  show the results of our analysis. The results displayed are only for $n=15$, as the figures for $n\in \{4,\ldots,14\}$ turned out to be very similar.
As Figure~\ref{fig3} shows, the behaviour of credit accumulation is similar for each choice of initial allocation~$y$.
Moreover, as expected, the average total deviation is clearly accumulating over time if arbitrary maximum matchings are chosen as solutions (see the left side figures of Figure~\ref{fig3}).
Under \emph{lexmin} and \emph{d1} there is still accumulation. However the credit system is indeed successfully mitigating against this effect, as the plots for \emph{lexmin+c} and \emph{d1+c} show (see the right side figures of Figure~\ref{fig3}). In particular, there is no indication that this behaviour will change if the number of rounds is larger than~$24$.

\subsection{Computational Time}\label{s-compreq}

We refer to Table~\ref{table2} for an overview of various computational times by our simulations.
We note that {\tt Lex-min} computes at most $n-1$ $d_i$-values, and in our experiments we actually found instances where $d_{n-1}$ was computed even for $n=10$.
However, Table~\ref{table2} shows that using lexicographically minimal maximum matchings instead of ones that only minimize the largest deviation $d_1$ from the target allocation does not require a significant amount of additional computation time. It can also be noticed from Table~\ref{table2} that, as expected, computing the Shapley value and the nucleolus is more expensive than computing the contribution value and the benefit value, especially as the number of countries $n$ is growing. Finally, we see from Table~\ref{table2} that the {\it game generation}, that is, computing the $2^n$ values $v(S)$, becomes by far the most expensive part when $n$ is growing.

\begin{table}[!ht]
\begin{center}
\setlength{\tabcolsep}{3.7pt}
\scriptsize
\begin{tabular}{|r|cccccccccccc|}
\hline
CPU times / Countries & 4 & 5 & 6 & 7 & 8 & 9 & 10 & 11 & 12 & 13 & 14 & 15 \\
\hline
Data preparation  & 0.59 & 0.64 & 0.56 & 0.53 & 0.53 & 0.52 & 0.56 & 0.54 & 0.50 & 0.52 & 0.51 & 0.50 \\
Graph building  & 20.74 & 20.81 & 20.82 & 20.69 & 20.73 & 20.67 & 20.77 & 20.70 & 20.57 & 20.67 & 20.59 & 20.66 \\
Game generation  & 0.06 & 0.10 & 0.18 & 0.34 & 0.66 & 1.29 & 2.59 & 5.06 & 9.90 & 19.96 & 39.17 & 79.58 \\
\quad \textit{ Shapley value }  & $\mathit{10^{-5}}$ & $\mathit{10^{-5}}$ & $\mathit{10^{-4}}$ & $\mathit{10^{-4}}$ & $\mathit{10^{-4}}$ & $\mathit{10^{-3}}$ & $\mathit{10^{-3}}$ & $\mathit{10^{-3}}$ & $\mathit{0.01}$ & $\mathit{0.02}$ & $\mathit{0.04}$ & $\mathit{0.08}$ \\
\quad \textit{ Nucleolus }  & $\mathit{10^{-3}}$ & $\mathit{10^{-3}}$ & $\mathit{10^{-3}}$ & $\mathit{10^{-3}}$ & $\mathit{0.01}$ & $\mathit{0.02}$ & $\mathit{0.03}$ & $\mathit{0.04}$ & $\mathit{0.06}$ & $\mathit{0.11}$ & $\mathit{0.22}$ & $\mathit{0.44}$ \\
\quad \textit{ Banzhaf }  & $\mathit{10^{-5}}$ & $\mathit{10^{-5}}$ & $\mathit{10^{-4}}$ & $\mathit{10^{-4}}$ & $\mathit{10^{-4}}$ & $\mathit{10^{-3}}$ & $\mathit{10^{-3}}$ & $\mathit{10^{-3}}$ & $\mathit{0.01}$ & $\mathit{0.02}$ & $\mathit{0.04}$ & $\mathit{0.08}$ \\
\quad \textit{ tau }  & $\mathit{10^{-5}}$ & $\mathit{10^{-5}}$ & $\mathit{10^{-4}}$ & $\mathit{10^{-4}}$ & $\mathit{10^{-4}}$ & $\mathit{10^{-3}}$ & $\mathit{10^{-3}}$ & $\mathit{10^{-3}}$ & $\mathit{0.01}$ & $\mathit{0.02}$ & $\mathit{0.04}$ & $\mathit{0.08}$ \\
\quad \textit{ Benefit }  & $\mathit{10^{-5}}$ & $\mathit{0.00}$ & $\mathit{10^{-5}}$ & $\mathit{10^{-6}}$ & $\mathit{10^{-5}}$ & $\mathit{0.00}$ & $\mathit{10^{-6}}$ & $\mathit{10^{-5}}$ & $\mathit{10^{-5}}$ & $\mathit{10^{-5}}$ & $\mathit{10^{-5}}$ & $\mathit{10^{-5}}$ \\
\quad \textit{ Contribution }  & $\mathit{10^{-5}}$ & $\mathit{0.00}$ & $\mathit{10^{-5}}$ & $\mathit{0.00}$ & $\mathit{10^{-6}}$ & $\mathit{0.00}$ & $\mathit{0.00}$ & $\mathit{10^{-5}}$ & $\mathit{10^{-5}}$ & $\mathit{10^{-5}}$ & $\mathit{0.00}$ & $\mathit{10^{-5}}$ \\ \hline
\textbf{ Total: lexmin+c }  & 21.75 & 21.95 & 21.98 & 22.00 & 22.38 & 22.96 & 24.46 & 26.82 & 31.51 & 41.73 & 60.93 & 101.46 \\ 
\textbf{ Total: d1+c }  & 21.74 & 21.92 & 21.92 & 21.90 & 22.26 & 22.80 & 24.28 & 26.64 & 31.33 & 41.56 & 60.71 & 101.32 \\
\textbf{ Total: lexmin }  & 21.78 & 21.96 & 21.98 & 21.99 & 22.36 & 22.93 & 24.42 & 26.78 & 31.46 & 41.68 & 60.81 & 101.4 \\
\textbf{ Total: d1 }  & 21.77 & 21.94 & 21.93 & 21.91 & 22.27 & 22.81 & 24.29 & 26.65 & 31.33 & 41.53 & 60.78 & 101.11 \\
\textbf{ Total: arbitrary }  & 21.28 & 21.55 & 21.56 & 21.57 & 21.94 & 22.51 & 24.00 & 26.41 & 31.13 & 41.35 & 60.51 & 101.18 \\ \hline
\end{tabular}
\vspace*{3mm}
\caption{Computational times for a single instance, broken down to the different computation tasks for \emph{lexmin+c}, while the total rows for the different scenarios are averaged over the four initial allocations.}
\label{table2}
\end{center}
\vspace*{-1.5cm}
\end{table}

\subsection{Coalitional Stability}

In order to assess the long-term coalitional stability of an IKEP, we turn our focus towards the core of the accumulated partitioned matching games. These games are obtained by summing up the 24 partitioned matching games of each of the 24 rounds of a simulation instance. That is, the {\it accumulated partitioned matching game} $(N,v)$ is obtained from the partitioned matching games $(N,v^h)$ on compatibility graphs $D^h$ for $h=1, \dots, 24$ by setting $v=\sum_{h=1}^{24} v^h$. We define the {\it accumulated initial allocation} as $y=\sum_{h=1}^{24} y^h$  and the {\it accumulated solution} as the 
accumulated number of kidney transplants $s=\sum_{h=1}^{24} s(M^h)$, where $M^h$ is the chosen matching in round~$h$.

All the accumulated partitioned matching games in our simulation had a nonempty core.
We evaluate how far away both the accumulated initial allocations and accumulated solutions are from violating a core allocation. We do this by taking the radius of the largest ball that can be fit into the core with it's center being an initial allocation, or a solution. Unsurprisingly, this radius is decreasing as the number of countries is increasing. Moreover, the distance of violating a core allocation is practically the same, independently of the chosen scenario. We refer to Tables \ref{table:initial} and~\ref{table:actual} in Appendix~\ref{a-core} for details and to Table~\ref{table:core_summary} for a summary obtained from these two tables by averaging over the number of countries for the \emph{lexmin+c} scenario.

From Table~\ref{table:core_summary} we see a high and similar level of stability for all choice of initial allocations. Although the Shapley and Banzhaf values provide consistently the smallest deviations (see Figure~\ref{fig4}), Table~\ref{table:core_summary} shows that the tau value (highest), the benefit value and the nucleolus provide higher levels of coalitional stability not only for the accumulated initial allocations, but also for the accumulated solutions.

\begin{table}[!ht]
\begin{center}
\setlength{\tabcolsep}{3.7pt}
\begin{tabular}{|r|ccccccc|}
\hline
 & Shapley  & nucleolus & Banzhaf  & tau & benefit  & contribution & Banzhaf*\\
\hline
Accumulated initial allocation & 50.46 & 53.34 & 50.08 & 53.62 & 53.40 & 48.15 & 49.74 \\ 
Accumulated solution & 50.39 & 53.10 & 50.03 & 53.43 & 53.19 & 48.13 & 48.39 \\ \hline
\end{tabular}
\vspace*{2mm}
\caption{Average distances, over $n$ ranging from $4$ to $15$, of accumulated initial allocations (first row) and accumulated solutions from violating a core inequality of the accumulated partitioned matching games under the \emph{lexmin+c} scenario. 
For example, by using the Shapley value as the initial allocation, every coalition of countries on average has at least 50.39 more kidney transplants by participating in the IKEP than they would be able to achieve on their own.}
\label{table:core_summary}
\end{center}
\vspace*{-1.5cm}
\end{table}

\begin{table}
\begin{center}
\setlength{\tabcolsep}{3.7pt}
\begin{tabular}{|r|rrr|}
\hline
\multirow{2}{*}{$n$} & not quasi- & \multirow{2}{*}{convex} & not convex,\\
& balanced & & tau $=$ benefit \\
\hline
4 & 0\% & 36.7\% & 55.3\% \\
5 & 0.0167\% & 9\% & 60\% \\
6 & 0\% & 2.2\% & 51.5\% \\
7 & 0\% & 0.9\% & 45\% \\
8 & 0\% & 0.4\% & 38.9\% \\
9 & 0.025\% & 0.3\% & 32.2\% \\
10 & 0.008\% & 0.1\% & 27.9\% \\
11 & 0.05\% & 0\% & 22.9\% \\
12 & 0.083\% & 0.1\% & 18.3\% \\
13 & 0.15\% & 0.1\% & 14.8\% \\
14 & 0.05\% & 0\% & 12.9\% \\
15 & 0.1\% & 0\% & 11.2\% \\ \hline
Total & 0.04\% & 4.14\% & 31.6\% \\ \hline
\end{tabular}
\vspace*{2mm}
\caption{The first columns refers to number of countries. The second, third and fourth column give, respectively, the percentage of non-quasibalanced games; percentage of convex games; and
 percentage of non-convex games with tau and benefit values for the initial allocations coinciding. The percentages of games are taken over all rounds, all initial allocations and all scenarios. The percentages of games in the last row are taken over all rounds, all initial allocations,  all scenarios and all numbers of countries.}
\label{table:tau*}
\end{center}
\vspace*{-1.5cm}
\end{table}

\subsection{Convexity and Quasibalancedness}\label{s-convex}

Recall that the tau value is only defined if the game is quasibalanced and that we replaced the tau value by the benefit value if the tau value is not defined. We also recall that tau value and benefit value coincide when the game is convex. 
Table~\ref{table:tau*} provides justification for this replacement.

\section{Conclusions}\label{s-con}

Our simulations showed that using maximum matchings that are  lexicographically minimal with respect to the country deviations from target allocations leads to a significant improvement for IKEPs. Moreover, they showed that this improvement is even more significant when the number of countries is large. This is relevant, as IKEPs, such as Eurotransplant, are under development and others, such as Scandiatransplant, are expected to grow. 

Both lexicographically minimal maximum matchings and maximum matchings that  only minimize the maximum deviation $d_1$ can be computed in polynomial time. In practice one might expect that the latter can still be computed faster. However, our simulations showed that computing them instead of maximum matchings that only minimize the maximum deviation indeed does not require any significant additional computational time (see Section~\ref{s-compreq}).

A challenging part of our project was to compute the nucleolus of partitioned matching games consisting of up to fifteen countries. For this we used the state-of-the-art \emph{Lexicographical Descent} method of \cite{BFN20}.

\medskip
\noindent
{\bf Future Research.}
All the above findings for $2$-way exchange cycles are also interesting to research for a setting with $\ell$-way exchange cycles for $\ell\geq 3$. The previous 
experimental studies~\cite{BGKPPV20,KNPV20} for $\ell=3$ only considered 3--4 countries.
To do meaningful experiments for a large number of countries, a new practical approach is required to deal with the computational hardness of computing optimal solutions (recall the aforementioned \NP-hard result of~\cite{Abr07} for the case where $\ell\geq 3$).

We also plan to consider directed compatibility graphs with {\it weights} $w(i,j)$ on the arcs $(i,j)$ representing the utility of transplant~$(i,j)$. Computing a maximum-weight solution that minimizes the weighted country deviation $d_1$ now becomes \NP-hard~\cite{BKPP19}. However, we could still consider the set of maximum-size solutions as our set ${\cal M}$ instead of the set of maximum-weight solutions. We can then find a maximum-weight matching that lexicographically minimizes the original country deviations $|x_p-s_p(M)|$. The main challenge is to set weights~$w(i,j)$ appropriately, since optimization policies may vary widely in national KEPs. In Europe, maximizing the number of transplants is the first objective (as in our setting). However, further scores are based on different objectives, such as improving the quality of the transplants, easing the complexity of the logistics or giving priority to highly sensitized patients; see~\cite{Biro_etal2021} for further details.

\bigskip
\noindent
{\it Competing Interests.} The authors declare that they have no known competing financial interests or personal relationships that could have appeared to influence the work reported in this paper.

\appendix

\section{Average Maximum Relative Deviations}\label{a-max}

\begin{figure}
\vspace*{-0.5cm}
		\resizebox{\textwidth}{!}{
			\includegraphics{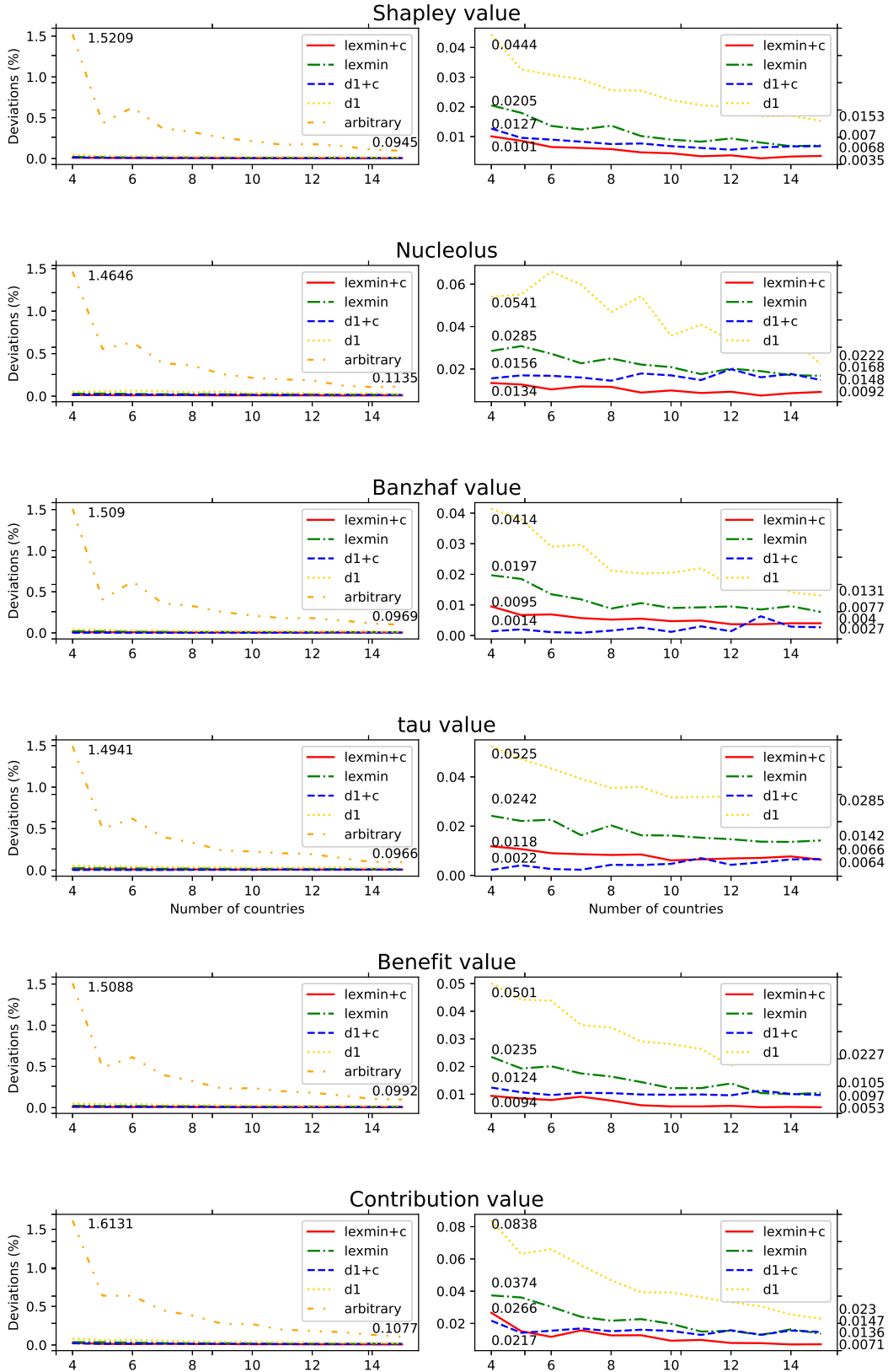}}
			\vspace*{-0.7cm}
			\caption{Average maximum relative deviations for the situation where all countries have the same size. The number of countries $n$ is ranging from $4$ to $15$. The figures on the right side zoom in on the figures from the left side by removing the results for the arbitrary matching scenario.}\label{fig5}
\end{figure}

\begin{figure}
\vspace*{-1cm}
		\resizebox{\textwidth}{!}{
			\includegraphics{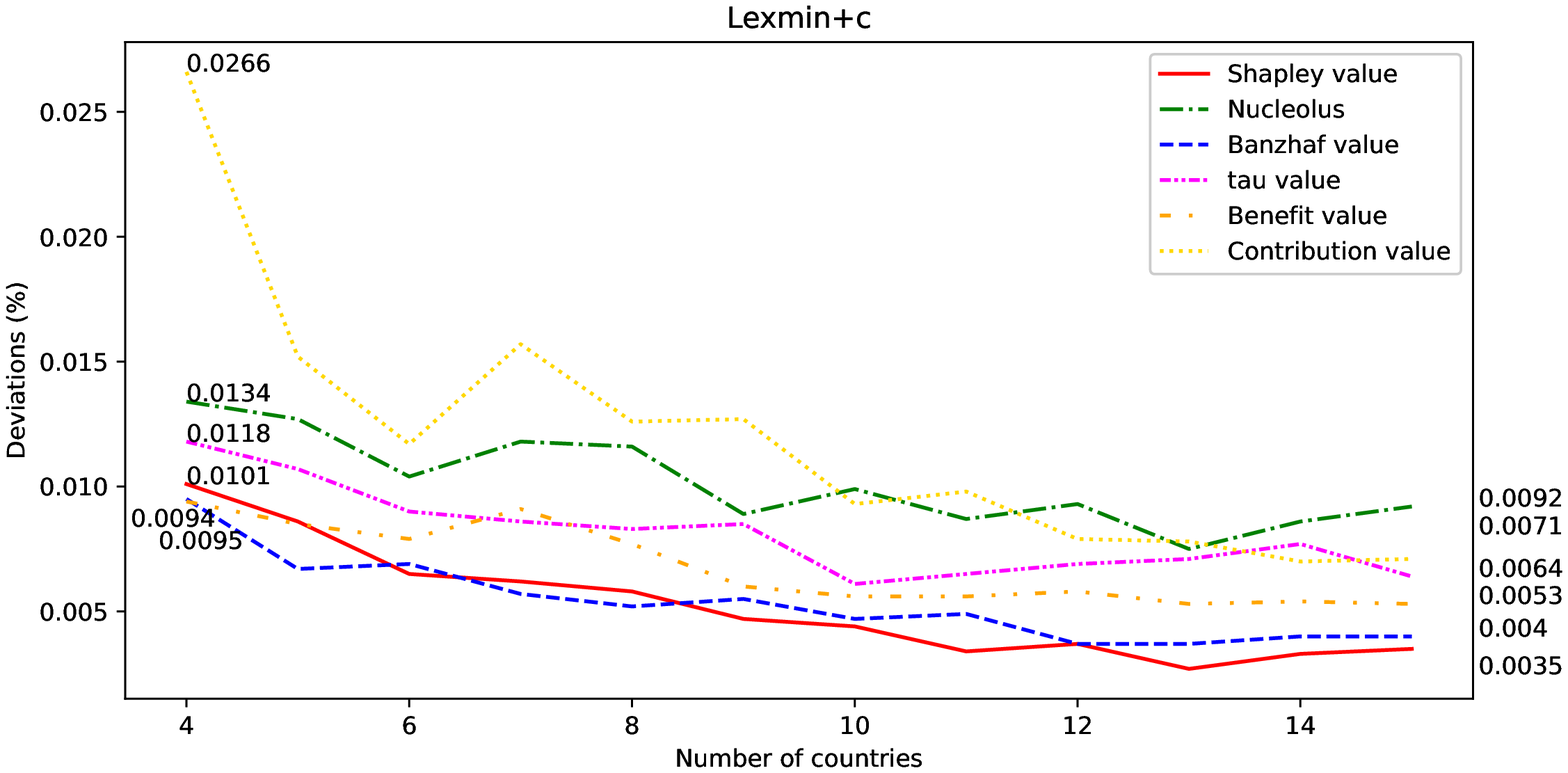}}
			\vspace*{-1.1cm}
			\caption{Displaying the four \emph{lexmin+c} graphs of Figure~\ref{fig5} in one plot.}\label{fig7}
			\vspace*{-0.8cm}
\end{figure}

\begin{figure}
		\resizebox{\textwidth}{!}{
			\includegraphics{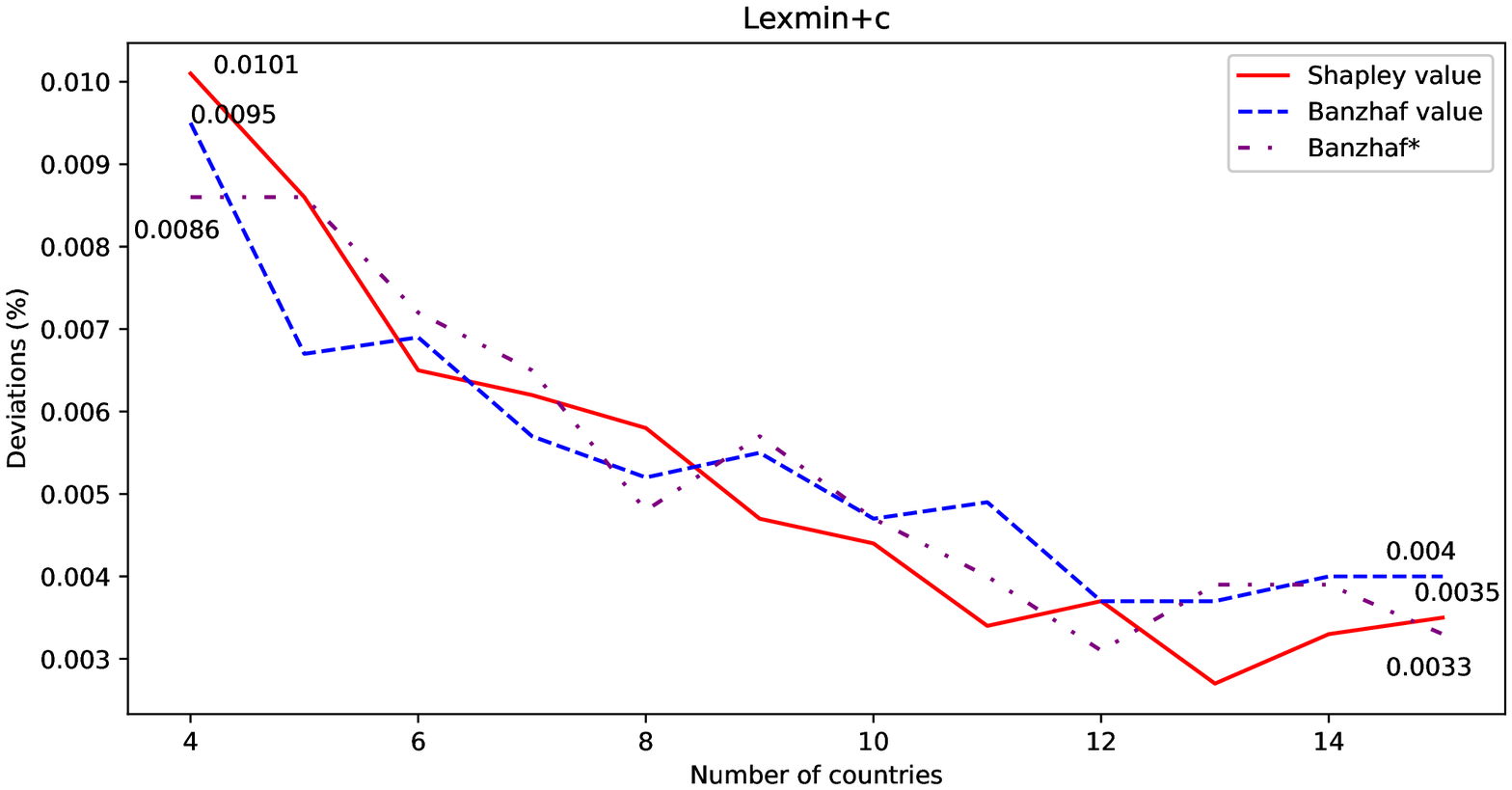}}
			\vspace*{-1.1cm}
			\caption{Comparing the \emph{lexmin+c} graphs for the Shapley value and Banzhaf value from Figure~\ref{fig5} with the one for the Banzhaf* value (same country sizes).}\label{fig7b}
\vspace*{-0.8cm}
\end{figure}

\begin{figure}
		\resizebox{\textwidth}{!}{
			\includegraphics{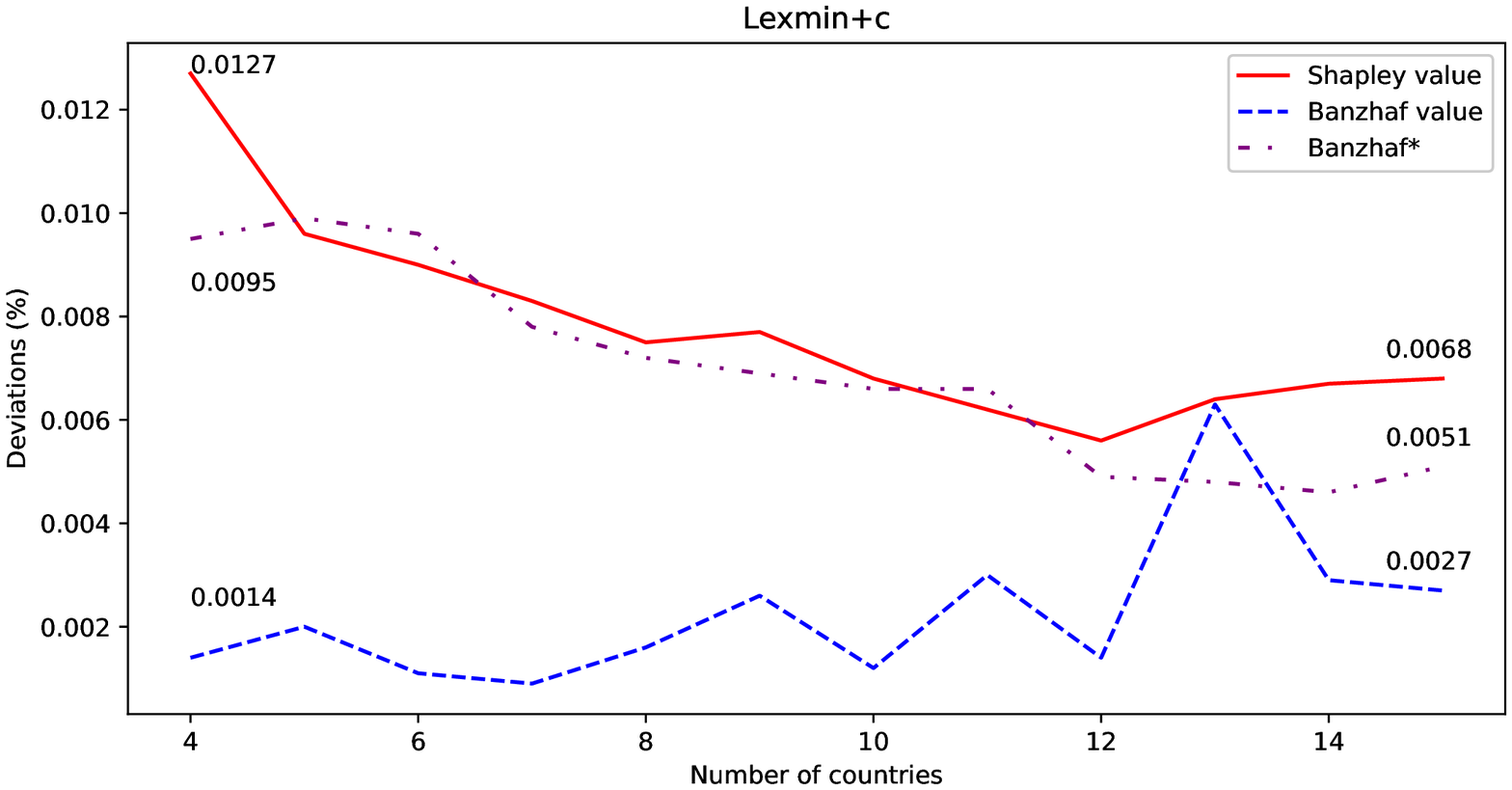}}
			\vspace*{-1.1cm}
			\caption{Comparing the \emph{d1+c} graphs for the Shapley value and Banzhaf value from Figure~\ref{fig5} with the one for the Banzhaf* value (same country sizes).}\label{fig7c}
\end{figure}

\begin{figure}
		\resizebox{\textwidth}{!}{
			\includegraphics{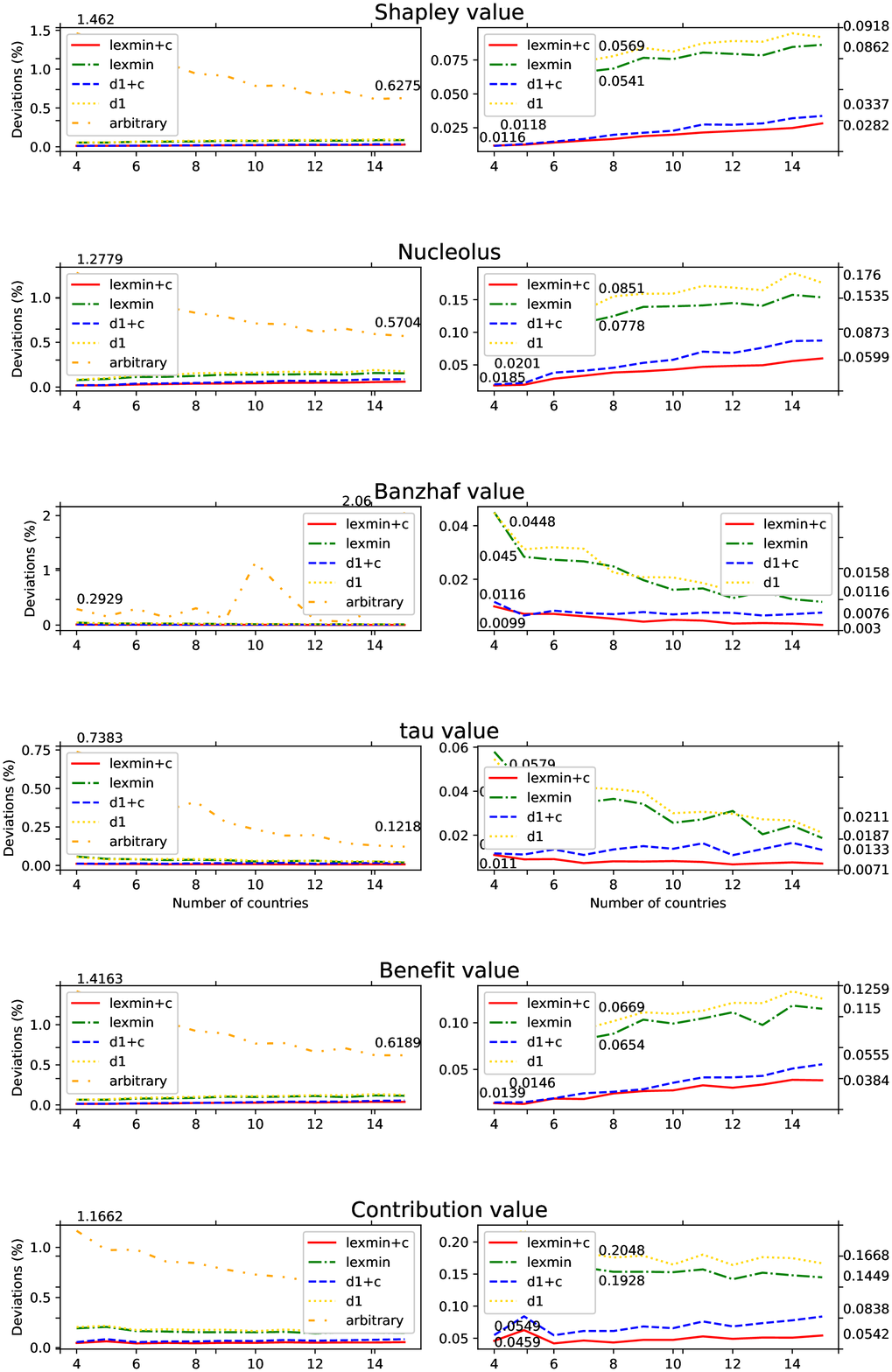}}
			\caption{Average maximum relative deviations for the situation where the countries vary in size. The number of countries $n$ is ranging from $4$ to $15$. The figures on the right side zoom in on the figures from the left side by removing the results for the arbitrary matching scenario.}\label{fig5var}
\end{figure}

\begin{figure}
\vspace*{-1cm}
		\resizebox{\textwidth}{!}{
			\includegraphics{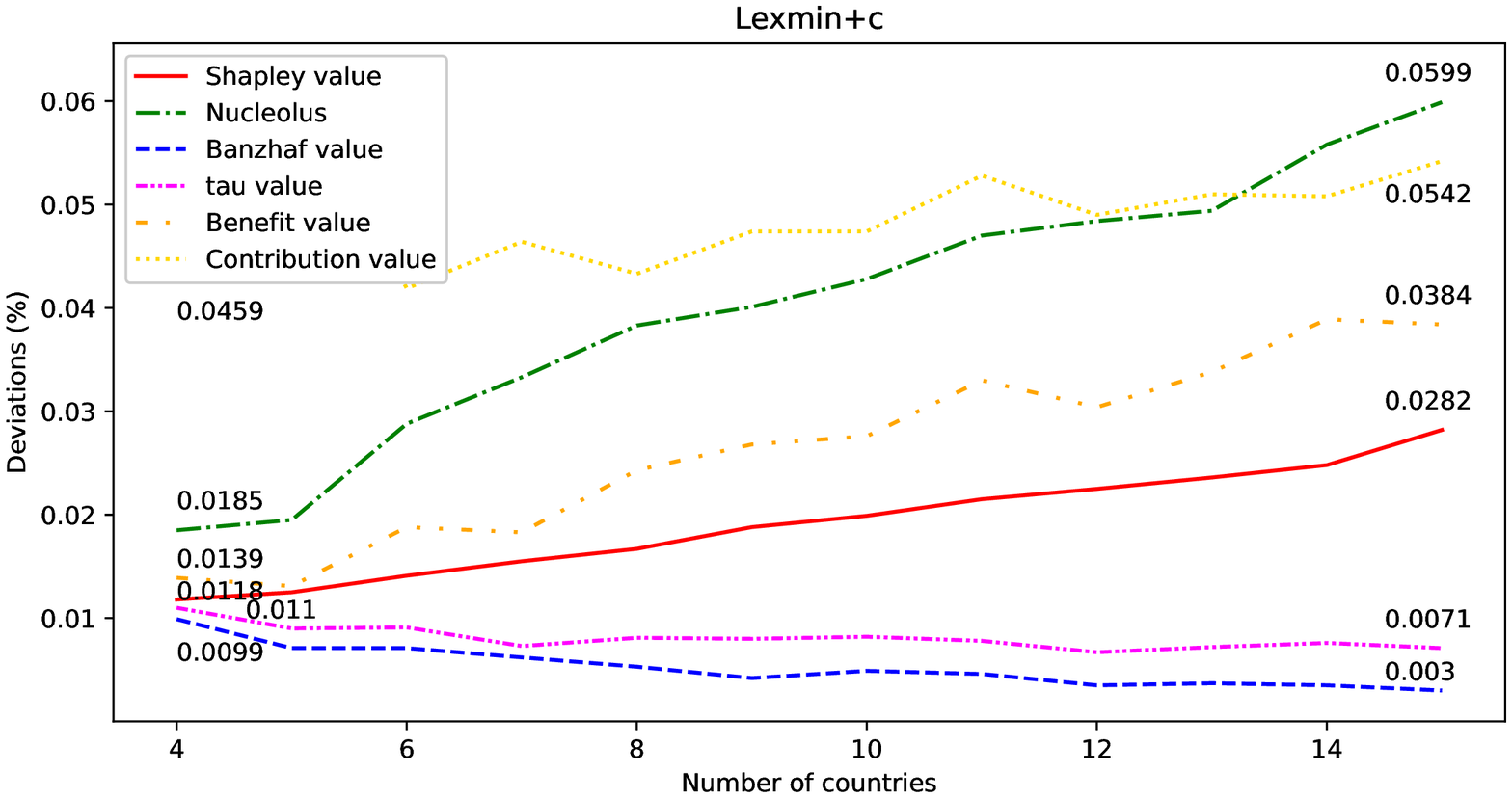}}
\vspace*{-1.1cm}
			\caption{Displaying the four lexmin+c graphs of Figure~\ref{fig5var} in one plot.}\label{fig7var} 
\vspace*{-0.8cm}
\end{figure}

\begin{figure}
		\resizebox{\textwidth}{!}{
			\includegraphics{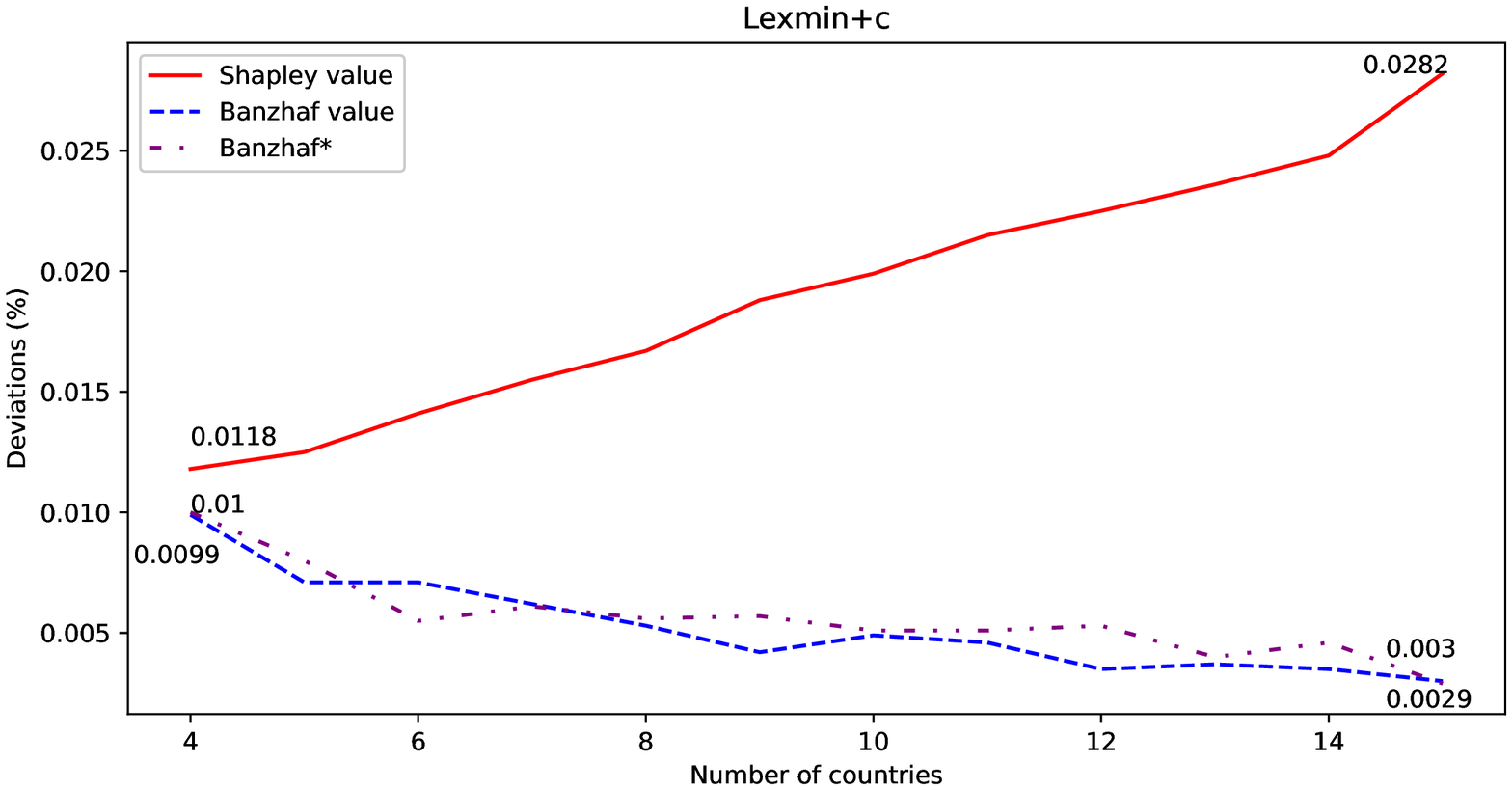}}
\vspace*{-1.1cm}
			\caption{Comparing the \emph{lexmin+c} graphs for the Shapley value and Banzhaf value from Figure~\ref{fig5var} with the one for the Banzhaf* value (varying country sizes).}\label{fig7bvar}
\vspace*{-0.8cm}
\end{figure}

\begin{figure}
		\resizebox{\textwidth}{!}{
			\includegraphics{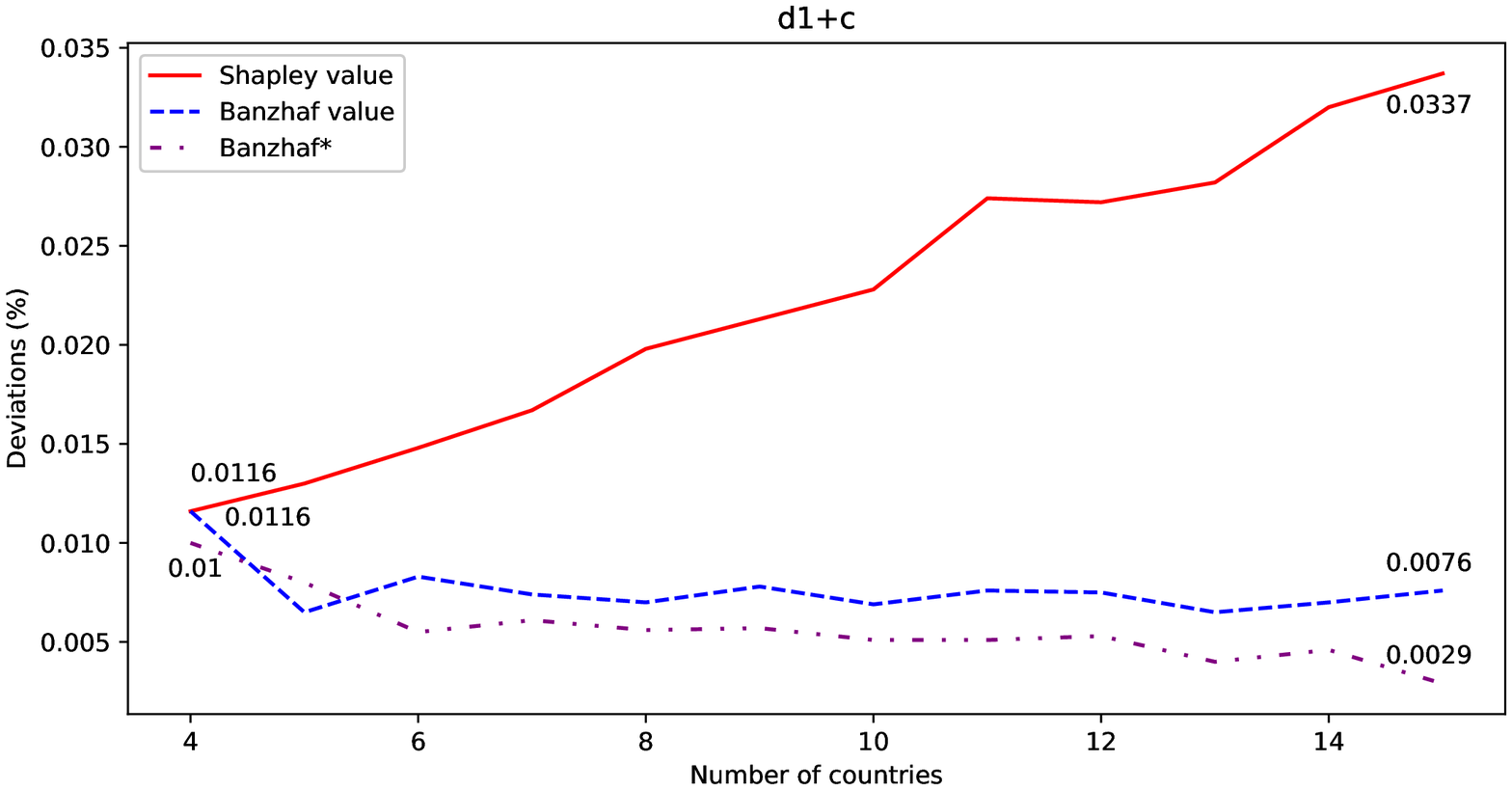}}
\vspace*{-1.1cm}
			\caption{Comparing the \emph{d1+c} graphs for the Shapley value and Banzhaf value from Figure~\ref{fig5var} with the one for the Banzhaf* value (varying country sizes).}\label{fig7cvar}
\end{figure}

We refer to Figures~\ref{fig5}--\ref{fig7cvar} as the counterparts of the figures in Section~\ref{s-results} where we use the average {\it maximum} relative deviation as evaluation measure instead of the average total relative deviation.

\section{Coalitional Stability}\label{a-core}

Tables~\ref{table:initial} and~\ref{table:actual}  show the average distances of accumulated initial allocations and accumulated number of transplants, respectively, from violating a core inequality.

\newpage

\newpage

\begin{table}
\begin{center}
\setlength{\tabcolsep}{3.7pt}
\scriptsize
\begin{tabular}{|r|ccccccc|}
\hline
Allocations \& policies & 4 & 5 & 6 & 7 & 8 & 9 & \\
\hline
\textbf{ Shapley: } &&&&&&&\\ 
lexmin+c & 105.25 & 85.64 & 70.31 & 59.65 & 51.39 & 44.60 & \\
lexmin & 106.08 & 85.29 & 69.95 & 59.25 & 51.54 & 44.70 & \\
d1+c & 106.12 & 85.84 & 69.93 & 59.29 & 52.06 & 45.21 & \\
d1 & 106.15 & 85.09 & 70.03 & 59.46 & 51.30 & 44.58 & \\ \hline
\textbf{ Nucleolus: } &&&&&&&\\ 
lexmin+c & 108.84 & 88.40 & 73.63 & 62.84 & 54.48 & 47.92 & \\
lexmin & 108.40 & 88.39 & 73.81 & 63.19 & 54.84 & 48.03 & \\
d1+c & 108.68 & 88.79 & 73.76 & 63.03 & 54.00 & 48.44 & \\
d1 & 108.34 & 88.33 & 73.96 & 63.52 & 54.57 & 47.86 & \\ \hline
\textbf{ Banzhaf: } &&&&&&&\\ 
lexmin+c & 105.08 & 84.88 & 68.95 & 59.31 & 50.22 & 43.86 & \\
lexmin & 105.38 & 84.75 & 69.68 & 58.83 & 50.50 & 43.49 & \\
d1+c & 104.72 & 83.75 & 69.43 & 58.87 & 50.78 & 43.74 & \\
d1 & 105.43 & 84.72 & 69.67 & 59.21 & 50.93 & 43.79 & \\
arbitrary & 105.20 & 84.46 & 68.22 & 58.13 & 50.59 & 43.04 & \\ \hline
\textbf{ tau: } &&&&&&&\\ 
lexmin+c & 107.78 & 87.52 & 72.96 & 61.84 & 54.17 & 48.15 & \\
lexmin & 107.21 & 87.48 & 73.17 & 62.18 & 54.29 & 48.01 & \\
d1+c & 107.34 & 87.61 & 72.84 & 62.15 & 54.21 & 48.06 & \\
d1 & 107.08 & 87.42 & 73.12 & 62.30 & 54.30 & 48.20 & \\
arbitrary & 106.72 & 86.75 & 71.83 & 60.87 & 53.90 & 46.79 & \\ \hline
\textbf{ Benefit: } &&&&&&&\\ 
lexmin+c & 107.45 & 87.47 & 73.18 & 62.11 & 54.62 & 48.21 & \\
lexmin & 107.43 & 87.32 & 72.52 & 61.57 & 54.11 & 47.93 & \\
d1+c & 107.37 & 87.55 & 72.60 & 61.90 & 54.67 & 47.87 & \\
d1 & 106.95 & 87.52 & 72.56 & 61.40 & 54.58 & 48.28 & \\ \hline
\textbf{ Contribution: } &&&&&&&\\ 
lexmin+c & 101.26 & 81.72 & 67.48 & 55.60 & 48.95 & 42.44 & \\
lexmin & 101.69 & 81.48 & 66.79 & 56.16 & 49.17 & 42.85 & \\
d1+c & 101.43 & 81.10 & 66.52 & 55.89 & 49.00 & 42.30 & \\
d1 & 101.98 & 81.12 & 66.53 & 56.31 & 49.40 & 42.66 & \\ \hline
\hline
Allocations \& policies & 10 & 11 & 12 & 13 & 14 & 15 & Total \\
\hline
\textbf{ Shapley: } &&&&&&&\\ 
lexmin+c & 40.43 & 35.65 & 32.31 & 28.78 & 26.86 & 24.71 & 50.46 \\
lexmin & 40.99 & 35.60 & 32.42 & 28.75 & 27.13 & 24.21 & 50.49 \\
d1+c & 40.04 & 35.44 & 32.19 & 28.72 & 26.69 & 24.13 & 50.47 \\
d1 & 40.57 & 36.01 & 32.01 & 29.07 & 26.76 & 24.35 & 50.45 \\ \hline
\textbf{ Nucleolus: } &&&&&&&\\ 
lexmin+c & 43.85 & 38.55 & 34.49 & 31.22 & 29.01 & 26.81 & 53.34 \\
lexmin & 43.43 & 38.52 & 34.37 & 31.34 & 29.31 & 26.64 & 53.36 \\
d1+c & 43.55 & 38.30 & 34.26 & 31.42 & 29.06 & 26.80 & 53.34 \\
d1 & 43.77 & 38.37 & 34.12 & 31.43 & 29.00 & 26.80 & 53.34 \\ \hline
\textbf{ Banzhaf: } &&&&&&&\\ 
lexmin+c & 39.70 & 35.09 & 30.64 & 27.49 & 26.05 & 23.53 & 50.08 \\
lexmin & 38.93 & 35.47 & 31.04 & 27.88 & 25.74 & 23.03 & 50.11 \\
d1+c & 39.07 & 34.94 & 31.53 & 27.54 & 26.02 & 23.14 & 49.98 \\
d1 & 39.62 & 35.08 & 30.87 & 27.58 & 26.03 & 23.15 & 50.21 \\
arbitrary & 39.63 & 34.27 & 30.25 & 27.42 & 25.72 & 23.11 & 49.73 \\ \hline
\textbf{ tau: } &&&&&&&\\ 
lexmin+c & 43.38 & 38.77 & 34.81 & 32.06 & 29.54 & 26.97 & 53.62 \\
lexmin & 43.53 & 38.62 & 34.73 & 31.89 & 29.61 & 27.21 & 53.63 \\
d1+c & 43.56 & 38.46 & 34.90 & 31.67 & 29.50 & 27.16 & 53.58 \\
d1 & 43.57 & 38.83 & 34.76 & 31.72 & 29.67 & 26.85 & 53.60 \\
arbitrary & 42.99 & 37.44 & 33.72 & 31.17 & 28.87 & 26.47 & 52.79 \\ \hline
\textbf{ Benefit: } &&&&&&&\\ 
lexmin+c & 43.66 & 38.87 & 35.25 & 32.07 & 30.03 & 27.92 & 53.40 \\
lexmin & 43.21 & 38.65 & 35.06 & 32.29 & 30.01 & 27.71 & 53.15 \\
d1+c & 43.36 & 38.83 & 35.19 & 32.46 & 29.86 & 27.79 & 53.29 \\
d1 & 43.13 & 38.92 & 34.95 & 32.30 & 29.81 & 27.71 & 53.18 \\ \hline
\textbf{ Contribution: } &&&&&&&\\ 
lexmin+c & 38.69 & 33.61 & 31.07 & 27.72 & 26.09 & 23.13 & 48.15 \\
lexmin & 38.47 & 33.69 & 30.90 & 28.04 & 25.88 & 23.43 & 48.21 \\
d1+c & 38.49 & 33.66 & 30.83 & 27.62 & 26.06 & 23.16 & 48.01 \\
d1 & 38.61 & 33.76 & 30.94 & 27.72 & 26.22 & 23.21 & 48.21 \\ \hline
\end{tabular}
\vspace*{2mm}
\caption{Average distance of accumulated initial allocations from violating a core inequality of the accumulated partitioned matching games for the four initial allocations, the four scenarios and the twelve country set sizes as well as total average over all country set sizes.}
\label{table:initial}
\end{center}
\end{table}

\begin{table}[!ht]
\begin{center}
\hspace*{-1cm}
\setlength{\tabcolsep}{3.7pt}
\scriptsize
\begin{tabular}{|r|ccccccc|}
\hline
Allocations \& policies & 4 & 5 & 6 & 7 & 8 & 9 & \\
\hline
\textbf{ Shapley: } &&&&&&&\\ 
lexmin+c & 105.24 & 85.58 & 70.22 & 59.57 & 51.37 & 44.55 & \\
lexmin & 105.79 & 84.85 & 69.60 & 58.51 & 50.77 & 43.77 & \\
d1+c & 106.17 & 85.84 & 69.98 & 59.32 & 51.98 & 45.06 & \\
d1 & 105.83 & 84.61 & 69.69 & 58.88 & 50.36 & 43.22 & \\
arbitrary & 47.92 & 35.98 & 29.64 & 22.49 & 19.99 & 14.50 & \\ \hline
\textbf{ Nucleolus: } &&&&&&&\\ 
lexmin+c & 108.75 & 88.27 & 73.37 & 62.59 & 54.28 & 47.60 & \\
lexmin & 107.19 & 86.75 & 72.14 & 61.53 & 52.39 & 46.01 & \\
d1+c & 108.64 & 88.66 & 73.72 & 62.80 & 53.76 & 48.08 & \\
d1 & 106.94 & 86.64 & 71.98 & 60.43 & 52.31 & 44.02 & \\
arbitrary & 47.92 & 35.98 & 29.64 & 22.49 & 19.99 & 14.50 & \\ \hline
\textbf{ Banzhaf: } &&&&&&&\\ 
lexmin+c & 105.02 & 84.82 & 68.91 & 59.24 & 50.16 & 43.8 & \\
lexmin & 105.01 & 84.43 & 69.20 & 58.21 & 49.89 & 42.78 & \\
d1+c & 104.82 & 83.82 & 69.33 & 58.85 & 50.76 & 43.61 & \\
d1 & 105.02 & 84.30 & 69.05 & 58.52 & 50.19 & 43.12 & \\
arbitrary & 47.92 & 35.98 & 29.65 & 22.51 & 20.09 & 14.81 & \\ \hline
\textbf{ tau: } &&&&&&&\\ 
lexmin+c & 107.7 & 87.55 & 72.8 & 61.59 & 54.02 & 47.76 & \\
lexmin & 106.62 & 86.35 & 71.97 & 60.87 & 52.81 & 46.09 & \\
d1+c & 107.35 & 87.51 & 72.77 & 62.03 & 53.87 & 47.82 & \\
d1 & 106.52 & 86.2 & 71.66 & 60.39 & 52.06 & 45.62 & \\
arbitrary & 47.92 & 35.98 & 29.65 & 22.51 & 20.09 & 14.81 & \\ \hline
\textbf{ Benefit: } &&&&&&&\\ 
lexmin+c & 107.48 & 87.39 & 73.02 & 61.88 & 54.39 & 47.99 & \\
lexmin & 107.03 & 86.29 & 71.51 & 60.26 & 52.36 & 46.17 & \\
d1+c & 107.32 & 87.50 & 72.55 & 61.76 & 54.53 & 47.52 & \\
d1 & 106.37 & 86.66 & 71.23 & 59.91 & 52.67 & 46.03 & \\
arbitrary & 47.92 & 35.98 & 29.64 & 22.49 & 19.99 & 14.50 & \\ \hline
\textbf{ Contribution: } &&&&&&&\\ 
lexmin+c & 101.15 & 81.79 & 67.55 & 55.62 & 48.85 & 42.31 & \\
lexmin & 102.06 & 82.25 & 66.94 & 56.30 & 49.76 & 42.49 & \\
d1+c & 101.57 & 81.20 & 66.79 & 55.84 & 49.25 & 42.33 & \\
d1 & 102.38 & 81.24 & 66.72 & 56.17 & 49.50 & 42.19 & \\
arbitrary & 47.92 & 35.98 & 29.64 & 22.49 & 19.99 & 14.50 & \\ \hline
\hline
Allocations \& policies & 10 & 11 & 12 & 13 & 14 & 15 & Total \\
\hline
\textbf{ Shapley: } &&&&&&&\\ 
lexmin+c & 40.39 & 35.55 & 32.19 & 28.74 & 26.68 & 24.60 & 50.39 \\
lexmin & 39.89 & 34.58 & 31.43 & 27.68 & 25.99 & 23.10 & 49.66 \\
d1+c & 39.90 & 35.33 & 32.08 & 28.58 & 26.48 & 24.07 & 50.40 \\
d1 & 39.53 & 34.76 & 30.70 & 27.50 & 25.27 & 22.93 & 49.44 \\
arbitrary & 13.81 & 11.19 & 8.11 & 7.23 & 7.52 & 6.79 & 18.76 \\ \hline
\textbf{ Nucleolus: } &&&&&&&\\ 
lexmin+c & 43.65 & 38.22 & 34.28 & 31.06 & 28.79 & 26.36 & 53.10 \\
lexmin & 41.46 & 36.29 & 32.43 & 28.96 & 27.07 & 24.37 & 51.38 \\
d1+c & 43.01 & 37.65 & 33.48 & 30.56 & 28.13 & 25.69 & 52.85 \\
d1 & 40.48 & 34.94 & 30.74 & 27.60 & 25.14 & 23.40 & 50.38 \\
arbitrary & 13.81 & 11.19 & 8.11 & 7.23 & 7.52 & 6.79 & 18.76 \\ \hline
\textbf{ Banzhaf: } &&&&&&&\\ 
lexmin+c & 39.66 & 35.04 & 30.57 & 27.45 & 25.99 & 23.49 & 50.03 \\
lexmin & 38.17 & 34.96 & 30.43 & 27.03 & 24.84 & 22.4 & 49.51\\
d1+c & 38.99 & 34.94 & 31.43 & 27.35 & 25.89 & 23.15 & 49.94 \\
d1 & 38.64 & 34.02 & 29.8 & 26.24 & 24.74 & 22.17 & 49.34 \\
arbitrary & 14.47 & 11.46 & 8.80 & 8.00 & 7.96 & 7.29 & 19.47 \\ \hline
\textbf{ tau: } &&&&&&&\\ 
lexmin+c & 43.3 & 38.53 & 34.47 & 31.84 & 29.21 & 26.73 & 53.43 \\
lexmin & 41.65 & 36.61 & 32.96 & 29.66 & 27.6 & 25 & 52.01\\
d1+c & 42.91 & 37.85 & 34.14 & 30.64 & 28.76 & 26.24 & 53.10 \\
d1 & 40.47 & 35.39 & 31.73 & 28.31 & 26.29 & 22.99 & 51.09 \\
arbitrary & 14.47 & 11.46 & 8.80 & 8.00 & 7.96 & 7.29 & 19.47 \\ \hline
\textbf{ Benefit: } &&&&&&&\\ 
lexmin+c & 43.45 & 38.59 & 34.94 & 31.82 & 29.69 & 27.58 & 53.19 \\
lexmin & 41.22 & 36.78 & 32.63 & 30.09 & 27.68 & 25.14 & 51.43 \\
d1+c & 42.91 & 38.54 & 34.56 & 31.85 & 29.26 & 26.85 & 52.93 \\
d1 & 40.33 & 36.12 & 31.98 & 28.82 & 26.68 & 24.39 & 50.93 \\
arbitrary & 13.81 & 11.19 & 8.11 & 7.23 & 7.52 & 6.79 & 18.76 \\ \hline
\textbf{ Contribution: } &&&&&&&\\ 
lexmin+c & 38.71 & 33.73 & 31.11 & 27.58 & 25.98 & 23.18 & 48.13 \\
lexmin & 38.57 & 33.79 & 30.45 & 27.63 & 25.57 & 23.24 & 48.25 \\
d1+c & 38.40 & 33.48 & 30.41 & 27.44 & 25.95 & 23.15 & 47.98 \\
d1 & 37.57 & 33.59 & 29.87 & 26.90 & 24.57 & 22.33 & 47.75 \\
arbitrary & 13.81 & 11.19 & 8.11 & 7.23 & 7.52 & 6.79 & 18.76 \\ \hline
\end{tabular}
\vspace*{2mm}
\caption{Average distance of accumulated number of transplants from violating a core inequality of the accumulated partitioned matching games for the four initial allocations, the four scenarios and the sanity check of arbitrary matching, for the 12 country set sizes as well as total average over all country set sizes.}
\label{table:actual}
\end{center}
\end{table}

\end{document}